\documentclass[12pt]{article}
\def\slash#1{\setbox0=\hbox{$#1$}#1\hskip-\wd0\hbox to\wd0{\hss\sl/\/\hss}}
\usepackage{epsf, cite, amsmath, amssymb}
\usepackage{epsfig}
\usepackage[all]{xy}
\usepackage{dsfont}
\usepackage[table]{xcolor}
\usepackage{float}

\numberwithin{equation}{section}
\usepackage{subfigure}
\usepackage{graphicx}

\makeatletter
\renewcommand\section{\@startsection {section}{1}{\z@}%
                                   {-3.5ex \@plus -1ex \@minus -.2ex}
                                    {2.3ex \@plus.2ex}%
                                   {\normalfont\large\bfseries}}
\renewcommand\subsection{\@startsection{subsection}{2}{\z@}%
                                     {-3.25ex\@plus -1ex \@minus -.2ex}%
                                     {1.5ex \@plus .2ex}%
                                     {\normalfont\bfseries}}
\makeatother

\newcommand{\bea}{\begin{eqnarray}}
\newcommand{\eea}{\end{eqnarray}}
\newcommand{\be}{\begin{equation}}
\newcommand{\ee}{\end{equation}}

\def\pref#1{(\ref{#1})}

\begin{document}

\begin{titlepage}

\begin{center}



\vskip 2 cm
{\Large \bf Crossing, Modular Averages and $N \leftrightarrow k $  in WZW Models}\\
\vskip 1.25 cm { Ratul Mahanta
and Anshuman Maharana }

{\vskip 0.75cm
Harish-Chandra Research Institute, \\
HBNI, Chhatnag Road, Jhunsi,
Allahabad -  211019, India.}

\end{center}

\vskip 2 cm

\begin{abstract}
\baselineskip=18pt

We consider the construction of genus zero correlators of $SU(N)_k$ WZW models involving two Kac Moody primaries in the fundamental and two in the anti-fundamental representation from modular averaging of the contribution of the vacuum conformal block. In cases where we find the orbit of the vacuum conformal
block to be finite, modular averaging  reproduces the exact result for the correlators. In other cases, we perform the modular averaging numerically, the results are in agreement with the exact answers. We find a close relationship between the modular averaging sums of the theories
related by level rank duality.  We establish a one to one correspondence
between elements of the orbits of the vacuum conformal blocks of dual theories. The  contributions of paired terms to their respective correlators  are simply related.  One consequence of this is that the ratio between the OPE coefficients associated with dual correlators can be obtained analytically without  performing the sums involved in the modular averagings. The pairing of terms in the modular averaging sums for dual theories suggests  an interesting connection between level rank duality and semi-classical  holographic computations of the correlators in the theories.

\let\thefootnote\relax\footnotetext{electronic address: {\tt {ratulmahanta@hri.res.in, anshumanmaharana@hri.res.in} }}

\end{abstract}

\end{titlepage}

\pagestyle{plain}
\setcounter{page}{1}
\newcounter{bean}
\baselineskip18pt

\tableofcontents

\section{Introduction}

  The bootstrap  \cite{Ferrara:1973yt, Polyakov:1974gs} serves as an extremely useful tool in the study of conformal field theories (see \cite{Poland:2018epd, Simmons-Duffin:2016gjk, Rychkov:2016iqz, Paulos:2014vya} for reviews). An interesting direction of study  is its interplay with duality symmetries. For example,  in  \cite{S}  it was found that S-duality invariant points of N=4 supersymmetric Yang-Mill saturate the bootstrap bounds on the anomalous dimensions of low twist non-BPS operators, in \cite{3c} it was found that crossing has interesting implications for  the structure of the S-matrix in Chern Simons theories with matter.  Recently, a rather simple proposal has been put forward to generate crossing symmetric genus zero correlation functions
 in two dimensional conformal field theories \cite{Maloney:2017}. In this paper, we construct correlation functions in $SU(N)_k$ WZW models using the proposal and examine level rank duality of the models in this context.

     In two dimensions, crossing  together with modular invariance has provided strong constraints from the early days \cite{BPZ:1984, KZ:1984, Cappelli:1986hf, Cardy:1986ie, Gepner:1986wi,  Verlinde:1988sn, Dijkgraaf:1988tf, Moore:1988uz, Moore:1988ss, Bouwknegt:1992wg}. For some recent developments in 2D bootstrap see
\cite{Hellerman:2009bu} - \cite{Anton:2019}, and in particular  \cite{Collier:2016cls} - \cite{Bae:2018qym} for work on theories with currents. 
    The basic idea in \cite{Maloney:2017} is to make use of  transformation properties of  conformal blocks under crossing  to arrive at crossing symmetric candidate correlation functions.  Correlation functions are generated by starting from a seed
 contribution (as given by the contributions of conformal blocks of  some primaries of low dimension running in the intermediate channel) and summing over the orbit of the seed under crossing transformations to obtain a crossing symmetric candidate correlation function. In two dimensions, crossing symmetry acts as the modular group on conformal  blocks. Thus the sum over the orbit of the seed contribution corresponds to ``modular averaging" \footnote{This is very similar in spirit to the proposal of
 \cite{Castro:2012} to compute partition functions from  vacuum characters.}.
 It was shown in \cite{Maloney:2017}  that  modular averaging  can be used to successfully  compute genus zero four point functions of minimal models.  Modular averaging has appeared in the physics literature in the context of three-dimensional quantum gravity and is often referred to as Farey tail sums (see e.g.~\cite{Dijkgraaf:2000, Kraus:2008, Murthy:2009dq, Maloney:2010, Duncan:2011, Castro:2011, Keller:2015}).   It was argued in \cite{Maloney:2017} that terms that arise from the orbit of the seed contribution
would arise naturally in a semiclassical holographic $AdS_{3}$ dual computation of the CFT correlator.

     Our focus will be on WZW correlators of \cite{KZ:1984}, involving two Kac-Moody primaries in the fundamental and two in the anti-fundamental representation. We find
that the correlators can be constructed from modular averaging of the contribution of the vacuum block.  Primary examples of models where the sums can be done exactly are models with $N=k$ (the orbits for these models are finite).  For models where we have not been able to show that the orbit is finite, we consider examples with specific values of $N$ and $k$, and  perform the averaging numerically. 


     An interesting feature of WZW models is level rank duality \cite{Schnitzer1}. Dual primary fields
under $N \leftrightarrow k$ are related by transposition of the Young tableaux of their representations. The correlators considered in this paper are the simplest related 
to each other by this duality.   From the point of view of modular averaging, both $N$ and $k$ simply appear   as parameters in the matrices  associated with the action of the modular group on the conformal blocks. Thus modular averaging puts $N$ and $k$  in a more equal footing; one can hope that   writing correlators as modular averages can  reveal various  aspects of level rank duality. This expectation is borne out. We establish a one to one correspondence
between elements of the orbits of the vacuum conformal blocks of dual theories. The  contributions of paired terms to their respective correlators  are simply related.  This allows us to obtain the ratio between the OPE coefficients associated with dual correlators analytically without  performing the sums involved in the modular averagings. The pairing of terms also indicates that holographic
computations can make some properties of the level rank duality manifest.

This paper is organised as follows. In section \ref{review}, we briefly review some basic ingredients that will be necessary for our analysis. In section
\ref{blocks} (and Appendix \ref{apblocks}) we obtain the transformation properties of the conformal blocks of the correlators under the action of the modular group.
In section \ref{Saverage} (and Appendix \ref{exnneqk}, \ref{apmodav}) we compute correlators by modular averaging. In section \ref{Snk}, we examine level rank duality. 


\section{Review}
\label{review}

We start by recalling some basic facts about
four point functions in two dimensional conformal field theories. We then go on to describe  the proposal of  \cite{Maloney:2017}
to construct crossing symmetric correlation functions from modular averaging.

The four-point correlator of operators $O_1$, $O_2$, $O_3$ and $O_4$ in 2D CFTs on the Riemann sphere can be written as the product of
a factor that determines its transformation properties under global conformal transformations and a function of a conformally invariant cross ratio. It will be our convention
to take
\begin{equation}
\label{cor}
  \langle  O_1(z_{1},\bar{z}_{1})O_2(z_{2},\bar{z}_{2})O_3(z_{3},\bar{z}_{3})O_4(z_{4},\bar{z}_{4}) \rangle =   G_0\big(z_{a},\bar{z}_{a}\big) G_{1234}(x,\bar{x}) \end{equation}
with
\begin{equation}
\label{prefac}
  G_0\big(z_{a},\bar{z}_{a}\big) = \prod_{a<b} \big(z_{ab}^{\mu_{ab}}\cdot \bar{z}_{ab}^{\bar{\mu}_{ab}}\big), 
\end{equation}
where $z_{ab}=z_a-z_b$  $(a,b = 1 . \  .  4)$, $\mu_{ab}=(\frac{1}{3}\sum_{c=1}^4h_c)-h_a-h_b$  ($h_{i}$ being the dimensions of the operators $O_{i}$)
and the cross ratio
\begin{equation}
\label{crossdef}
x=\frac{z_{12}z_{34}}{z_{14}z_{32}}.
\end{equation}
 Conformal transformations can be used to set $z_2$ to $0$ and $z_3$ to $1$ and set $z_4$ to infinity, the coordinate $z_1$ then corresponds to
 the cross ratio. Thus the cross ratio space is the Riemann sphere with three punctures.

  Correlators in two dimensional CFTs can be constructed from holomorphic and antiholomorophic conformal blocks.
Although correlators need to be single valued functions of the cross ratio space\footnote{We will be dealing with bosonic operators.}, there is no such requirement on the conformal blocks. Conformal blocks have
monodromies in the cross ratio space. Thus it is natural to consider conformal blocks as functions in the universal covering space of the cross ratio space. This
is $\mathbb{H}_+ =\{u+iv \text{ }|\text{ }v>0 \text{ and } u,v\in\mathbb{R}\}$, the upper half plane\footnote{The observation  that conformal blocks should be  single-valued on the upper half plane was made in \cite{Zamolodchikov}, where an  elliptic recursion representation was obtained for them.}. The elliptic lambda function 
\begin{equation}
\label{eq:4}
  \lambda(\tau)=\left(\frac{\theta_2(\tau)}{\theta_3(\tau)}\right)^4 \text{, } 
\end{equation}
where $\tau = u + iv$ provides a surjective map ($x=\lambda(\tau)$) from $\mathbb{H}_{+}$ to the cross ratio space ~\cite{Weisstein1}. $PSL(2, \mathbb{Z})$ action on the upper half
plane has a  close connection to the map. Under the action of the generators of the modular group
\begin{equation}
 T:\tau \to \tau + 1  \phantom{a} \text{and}  \phantom{a} S:\tau \to -\frac{1}{\tau},
 \end{equation}
 images in the cross ratio space have rather simple transformations
 \begin{equation}
 \label{STact}
  T\cdot x=\frac{x}{x-1}  \phantom{a} \text{and}  \phantom{a} S\cdot x=1-x.
\end{equation}
Furthermore, the function $\lambda(\tau)$ is invariant under the normal subgroup $\Gamma(2)$ of $PSL(2,\mathbb{Z})$: 
\begin{equation}
\lambda(\gamma\tau)=\lambda(\tau) \text{, } \forall\gamma\in\Gamma(2).
\end{equation}
Thus,  the condition that correlators have to be single valued in the cross ratio space translates to invariance under $\Gamma(2)$ in $\mathbb{H}_+$.

      At this stage, it is natural to seek for the interpretation of the action of the entire $PSL(2, \mathbb{Z})$ on the correlators in the CFT.   For this, one has to look
 at      crossing symmetry.  For a general ordering of the operators, we define
\begin{equation}
\label{corgen}
  \langle  O_p(z_{p},\bar{z}_{p})O_q(z_{q},\bar{z}_{q})O_r(z_{r},\bar{z}_{r})O_s(z_{s},\bar{z}_{s}) \rangle =   G_0\big(z_{a},\bar{z}_{a}\big) G_{pqrs}(x_{pqrs},\bar{x}_{pqrs}), \end{equation}
with $G_{0}$ as defined in \pref{prefac} and
\begin{equation}
\label{gencross}
     x_{pqrs} = \frac{z_{pq} z_{rs}}{z_{ps} z_{rq}}.
\end{equation}
Note that with this we have  $x = x_{1234}$, where $x$ is the cross ratio introduced in \pref{crossdef}. Our choice of $G_0$ is invariant under permutations of $z_{a}$
thus  crossing symmetry reduces to the statement that $G_{abcd}(x_{abcd})$ is invariant under action of the same permutation on $\{a, b, c, d\}$ in both the subscripts. Permutations that leave the cross
ratio $x$ invariant yield:
\begin{equation}
\label{triv}
  G_{1234} (x, \bar{x}) = G_{2143} (x, \bar{x}) = G_{3412} (x, \bar{x}) = G_{4321} (x, \bar{x}).
\end{equation}
On the other hand, permutations which act non-trivially on the cross ratio\footnote{These relations differ from the ones  in~\cite{Maloney:2017} since our  choice for the cross-ratio $x$ is different.} give
\begin{equation}
\label{nontri}
  \begin{aligned}
    G_{1234}(x,\bar{x})& = G_{1243}(\frac{x}{x-1},\frac{\bar{x}}{\bar{x}-1}) = G_{3241}(\frac{1}{1-x},\frac{1}{1-\bar{x}}) = G_{3214}(\frac{1}{x},\frac{1}{\bar{x}})\\
    & = G_{4231}(1-x,1-\bar{x}) = G_{4213}(\frac{x-1}{x},\frac{\bar{x}-1}{\bar{x}}).
  \end{aligned}
\end{equation}
 The arguments of the functions in \pref{nontri} can be related by the actions of $S$ and $T$ as given in \pref{STact}. The actions are isomorphic to the anharmonic group,  $S_3$. This is  precisely equal to $PSL(2, \mathbb{Z}) / \Gamma(2)$.   Thus crossing symmetry and single valuedness\footnote{Recall that  correlators need to be invariant under $\Gamma(2)$  so that they single valued.} together specify the full $PSL(2, \mathbb{Z})$ action on the correlators.
Combining  \pref{STact},\pref{triv} and \pref{nontri} they can be written in a very compact form \cite{Maloney:2017}:
 \begin{equation}
\label{master}
  \vec{G}(\gamma\tau,\gamma\bar{\tau})= \sigma(\gamma) \cdot \vec{G}(\tau,\bar{\tau}) \text{, }  \phantom{ab} \gamma \in PSL(2,\mathbb{Z})
\end{equation}
 where
 \begin{equation}
 \label{gvec}
     \vec{G} = ( G_{1234}(\tau, \bar{\tau}), G_{2134}(\tau, \bar{\tau}), G_{4132}(\tau, \bar{\tau}), G_{1432}(\tau, \bar{\tau}), G_{2431}(\tau, \bar{\tau}), G_{4231}(\tau, \bar{\tau}))^{t}     
   \end{equation}
 and  $\sigma(\gamma)$ are the six dimensional matrices associated with the linear representation of $PSL(2, \mathbb{Z})/  \Gamma(2) = S_3$ with
 \begin{equation}
 \sigma(S)=
  \begin{pmatrix}
    0 & 0 & 0 & 0 & 0 & 1\\
    0 & 0 & 1 & 0 & 0 & 0\\
    0 & 1 & 0 & 0 & 0 & 0\\
    0 & 0 & 0 & 0 & 1 & 0\\
    0 & 0 & 0 & 1 & 0 & 0\\
    1 & 0 & 0 & 0 & 0 & 0
  \end{pmatrix}
  \phantom{a} \textrm{and} \phantom{b}
\sigma(T)=
  \begin{pmatrix}
    0 & 1 & 0 & 0 & 0 & 0\\
    1 & 0 & 0 & 0 & 0 & 0\\
    0 & 0 & 0 & 1 & 0 & 0\\
    0 & 0 & 1 & 0 & 0 & 0\\
    0 & 0 & 0 & 0 & 0 & 1\\
    0 & 0 & 0 & 0 & 1 & 0
  \end{pmatrix}.
\end{equation}
We note that there is further simplification when all or some of the operators $O_{a}$ are identical. For instance, in the case that all the four operators are identical 
$\vec{G}$ has only one independent component. Equation \pref{master} requires it to be a modular invariant scalar.

     Modular averaging can be used to obtain solutions of equations of the form of \pref{master}. The general structure
 of four point functions in a CFT gives  fiducial  functions   over which the averaging can be performed. 
Conformal invariance implies that the stripped correlators in \pref{corgen} can be written as a sum over contributions
associated with conformal primaries ($\phi_k$):
\begin{equation}
\label{gencor1}
  G_{pqrs}(y,\bar{y})=\sum_{k}C_{O_p O_q \phi_k} C_{O_r O_s \phi_k} \times y^{h_{\phi_k} - \frac{\mathfrak{H}}{3}} \bar{y}^{\bar{h}_{\phi_k} - \frac{\bar{\mathfrak{H}}}{3}} F^{\phi_k}_{pqrs}(y,\bar{y}),
\end{equation}
where $C_{O_a O_b \phi_k}$, $C_{O_c O_d \phi_k}$ are three point structure constants, 
${\mathfrak{H}=(h_a+h_b+h_c+h_d)}$ and ${\bar{\mathfrak{H}}=(\bar{h}_a+\bar{h}_b+\bar{h}_c+\bar{h}_d)}$. The
functions $F^{\phi_k}_{pqrs}(y,\bar{y})$ are analytic at $y, \bar{y} =0$ and $F^{\phi_k}_{pqrs}(0,0) = 1$. It will be
 our convention to call $\{ y^{h_{\phi_k} - \frac{\mathfrak{H}}{3}} \bar{y}^{\bar{h}_{\phi_k}-\frac{\bar{\mathfrak{H}}}{3}} 
 F^{\phi}_{pqrs}(y,\bar{y}) \}$ as  the conformal block corresponding to primary $\phi_k$. These can be further
 factorized into holomorphic and anti-holomorchic conformal blocks for each $\phi_k$.  Given the form
 of \pref{gencor1}, in the limit of $y \to 0$ the stripped correlator is well approximated by including contributions from the
 low lying primaries that appear in the sum i.e.  
\begin{equation}
\label{gencor2}
  G_{pqrs}(y,\bar{y})  \approx G^{\rm light}_{pqrs}(y, \bar{y})=\sum_{k \leq k_{\rm max}}C_{O_p O_q \phi_k} C_{O_r O_s \phi_k} \times y^{h_{\phi_k} - \frac{\mathfrak{H}}{3}} \bar{y}^{\bar{h}_{\phi_k} - \frac{\bar{\mathfrak{H}}}{3}} F^{\phi_k}_{pqrs}(y,\bar{y})
  \phantom{ab} \textrm{for}   \phantom{ab} y \to 0.
\end{equation}
where the sum now runs over primaries which have weights less than or equal to 
$(h_{k_{\rm max}}, \bar{h}_{k_{\rm max}})$. The simplest approximation is to keep only the primary with the lowest
weight.   Reference \cite{Maloney:2017} proposed that modular averaging of $\vec{G}^{\rm light}$ can be used
to construct candidate CFT correlators which satisfy the requirements single-valuedness and crossing.
\begin{equation}
\label{aver}
  \vec{G}^{\rm candidate}(\tau,\bar{\tau})= \mathcal{N}^{-1} \cdot \sum_{\gamma \in PSL(2,\mathbb{Z})} \sigma^{-1}(\gamma) \cdot \vec{G}^{\rm light}(\gamma\tau,\gamma\bar{\tau}),
\end{equation}
where $\mathcal{N}$ is a normalisation which can be determined from the $\tau \to i\infty$ $(y \to 0)$ behaviour 
of $\vec{G}(\tau,\bar{\tau})$. In general, the sum in \pref{aver} is difficult to perform and might even need regularisation.
The complications associated with dealing with a sum involving vector valued modular objects can 
be ameliorated for correlators with identical operators. As described earlier, in the presence of identical operators,
 various components of $\vec{G}$ (as defined in \pref{gvec}) become related - the vector space effectively collapses
 to a lower dimensional one. As a result, the subgroup of $PSL(2, \mathbb{Z})$ that leaves any particular component of
 the vector inert under action of $\sigma(\gamma)$ is enhanced\footnote{In the case that all the operators a distinct, this
 subgroup is $\Gamma(2)$ for all the components}. If the subgroup associated with the component
 $G_{a}$ in the collapsed vector space is $\Gamma_{a}$, a natural candidate $G_{a}$ can be constructed by defining
 \begin{equation}
 \label{newav}
   G_{a}^{\rm candidate} (\tau, \bar{\tau})= \mathcal{N}^{-1} \cdot \sum_{\gamma \in \Gamma_{a}} {G}_{a}^{\rm light}(\gamma\tau,\gamma\bar{\tau}).
 \end{equation}

      The above program to obtain CFT correlators was implemented for minimal models in \cite{Maloney:2017}. It was 
 found that for a large number of them, the candidate correlators did match with the exact ones by taking only the 
 contribution of the Virasoro vacuum block while constructing $G^{\rm light}_{a}$ - the lightest block served the purpose.

\section{ $SU(N)_k$ WZW Model: Conformal Blocks, Actions of S and T}
\label{blocks}

    As mentioned in the introduction, our focus will be on WZW correlators  involving two Kac-Moody primaries in the fundamental and two in the anti-fundamental representation.  In this section, we will obtain the transformation properties of the conformal blocks  associated with the correlators  under the action of crossing.

 We begin by  recalling some basic facts about the correlators (our discussion follows that of \cite{KZ:1984, Gepner:1986wi, Francesco, Blumenhagen}) and in the process set up our notation.
 The $SU(N)$ WZW model at level $k$ on the two sphere is described by the action:
\begin{equation}
  \begin{aligned}
    S^{\rm WZW}_k[g] =  \frac{k }{16\pi}\int d^2z \textrm{ } {\rm{Tr}} (\partial^\mu g^{-1} \partial_\mu g)   - \frac{i k}{24\pi}\int_B d^3\vec{X} \textrm{ } \epsilon_{\alpha\beta\gamma} {\rm{Tr}} ({g}^{-1} \partial^\alpha {g} {g}^{-1} \partial^\beta {g} {g}^{-1} \partial^\gamma {g}), \phantom{and}  \\
    \hfill{k = 1,2, .. \phantom{abc}}
  \end{aligned}
\end{equation}
where $g(z, \bar{z})$ is a matrix valued bosonic field which takes values in the group $SU(N)$.  The second term 
is an integral over the three ball $B$, whose boundary is the two sphere. The pre-factors of the two terms in the 
action are chosen so that theory is conformal at the quantum level. The action enjoys an
$SU(N)(z) \times SU(N)(\bar{z})$ invariance. The associated currents are
\begin{equation}
\label{curdef}
    j(z)\equiv-k(\partial_z g)g^{-1},\text{ }\bar{j}(\bar{z})\equiv kg^{-1}(\partial_{\bar{z}} g)\\
\end{equation}
which can be expanded in terms of the generators of $SU(N)$ as
\begin{equation}
\label{curexpa}
    j(z)=\sum\nolimits_a j^a(z)t^a,\text{ }\bar{j}(\bar{z})=\sum\nolimits_a \bar{j}^a(\bar{z})t^a.
 \end{equation}
The Laurent series expansion coefficients of the currents  together with the Virasoro generators generate two copies of the Kac-Moody algebra at level $k$.

         Kac-Moody primaries serve as the highest weight states in the theory. For the $(N,k)$ theory the spectrum of Kac-Moody primaries 
consists operators transforming in all representations of $SU(N)$ which have integrable Young tableaux  i.e. those in which the number of columns is at most $k$.
The conformal dimension of a  Kac-Moody primary transforming in a representation $R$ is %
\begin{equation}
\label{dimfor}
    h_{R} =  \frac{C(R)}{2(k + N)},
\end{equation}       
where $C(R)$ is the quadratic Casimir of the representation.

 We will follow the notation
of \cite{KZ:1984} and denote a fundamental Kac-Moody primary by $g_{\alpha}^{\phantom{\alpha}\beta}(z, \bar{z})$, where $\alpha$ is a fundamental index of the $SU(N)$ left and $\beta$ is a fundamental index of the $SU(N)$ right. On the other hand, an anti-fundamental will be denoted by $g^{-1 \sigma}_{\rho}{}$, where 
where $\rho$ is an anti-fundamental index of the $SU(N)$ right and $\sigma$ is an anti-fundamental index of the $SU(N)$ left. The conformal dimension of these
fields can be easily obtained from \pref{dimfor}
\begin{equation}
\label{gdim}
       h_g= h_{g^{-1}} = \frac{N^2-1}{2N(k+N)}.
\end{equation}
For correlators involving two fundamentals and two anti-fundamentals, primaries that run in the intermediate channels will be as per the fusion rules
\begin{equation}
  g \times g^{-1}= \mathds{1}+ {\theta},  \phantom{+\cdots}\text{ }g \times g= \xi+\chi, \phantom{+\cdots} \text{ }g^{-1} \times g^{-1}= \xi+\chi,
\end{equation}
where $\mathds{1}$ is the identity field, $\theta$ the adjoint, $\xi$ the antisymmetric and $\chi$ the symmetric. The associated dimensions are
\begin{equation}
\label{fdim}
h_\mathds{1}=0, \phantom{ab} h_{{\theta}}=\frac{N}{N + k}, \phantom{ab} h_{\xi}=\frac{(N-2)(N+1)}{N(N + k)} \phantom{sb} \textrm{and}\phantom{sb} h_{\chi}=\frac{(N+2)(N-1)}{N (N + k)}.
\end{equation}

   Our main interest will be the correlator
\begin{equation}
\label{maincor}
 \langle gg^{-1}g^{-1}g \rangle \equiv \langle  {g_{\alpha_1}}^{\beta_1}(z_1,\bar{z}_1) \cdot {g^{-1}_{\beta_2}}^{\alpha_2}(z_2,\bar{z}_2) \cdot {g^{-1}_{\beta_3}}^{\alpha_3}(z_3,\bar{z}_3) \cdot {g_{\alpha_4}}^{\beta_4}(z_4,\bar{z}_4) \rangle
\end{equation} 
Recall that as per our conventions $\alpha_1, \alpha_4$ are $SU(N)$ left fundamental indices, $\alpha_2, \alpha_3$ are $SU(N)$ left anti-fundamental indices,   $\beta_1, \beta_4$ are $SU(N)$ right fundamental indices, $\beta_2, \beta_3$ are $SU(N)$ right anti-fundamental indices.
We will be eventually interested in making choices for the indices such that the correlator contains two pairs of identical operators so that we can carry out
modular averaging as per the prescription in \pref{newav}. For this we need the conformal blocks associated with the correlator and their transformations under the
modular group.

   The correlator has been studied in detail in \cite{KZ:1984}. We briefly describe their analysis adopting the discussion to our conventions. First, we define 
the stripped correlator $G^{\beta_1 \alpha_2 \alpha_3 \beta_4}_{\alpha_1 \beta_2 \beta_3 \alpha_4}(x,\bar{x})$ as in \pref{cor}
\begin{equation}
\label{strone}
\langle gg^{-1}g^{-1}g \rangle = \big(\prod_{a<b} z_{ab}^{\mu_{ab}} \bar{z}_{ab}^{\bar{\mu}_{ab}}\big)  G^{\beta_1 \alpha_2 \alpha_3 \beta_4}_{\alpha_1 \beta_2 \beta_3 \alpha_4}(x,\bar{x}), 
 \end{equation}
where $x$ is the cross ratio defined in \pref{crossdef}. Invariance of the correlator under $SU(N)$ left and right implies
\begin{equation}
\label{prod}
    G^{\beta_1 \alpha_2 \alpha_3 \beta_4}_{\alpha_1 \beta_2 \beta_3 \alpha_4}(x,\bar{x}) = \sum_{A,B=1,2} (I_A)(\bar{I}_B)G_{AB}(x,\bar{x}),
\end{equation}    
where 
 \begin{equation}  
 \label{tenstruc} 
    I_1=\delta_{\alpha_1}^{\alpha_2}\delta_{\alpha_4}^{\alpha_3},  \phantom{a} \bar{I}_1 = \delta_{\beta_2}^{\beta_1}\delta_{\beta_3}^{\beta_4}, \phantom{a} I_2 = \delta_{\alpha_1}^{\alpha_3}\delta_{\alpha_4}^{\alpha_2}  \phantom{a} \textrm{and} \phantom{a} \bar{I}_2 = \delta_{\beta_3}^{\beta_1}\delta_{\beta_2}^{\beta_4}.
\end{equation}
One then imposes the Knizhnik-Zamolodchikov (KZ) equations on the correlator. The KZ equations are a consequence of the Kac-Moody symmetries. For a correlator
involving Kac-Moody primaries $\phi_i$, transforming in the representations $R_{i}$ they are
\begin{equation}
\label{eq:8}
  \bigg[ \partial_{z_i}-\frac{1}{k+N}\sum_{j\neq i} \frac{\sum\nolimits_a t^a_{R_{i}}\otimes t^a_{R_{j}}}{z_i-z_j} \bigg] \langle\phi_1(z_1,\bar{z}_1)\cdots\phi_n(z_n,\bar{z}_n)\rangle =0, \text{ } \forall\textrm{ }i,
\end{equation}
where $t^{a}_{R_{i}} $ are $SU(N)$  generators in the representation $R_{i}$. Similar set of equations hold in the anti-holomorphic coordinates. Imposing
them on the correlator \pref{maincor}  yields the following equations for the matrix $G_{AB}$ defined in \pref{prod}.
\begin{equation}
  \frac{\partial G}{\partial x}=\bigg[\frac{1}{x}P+\frac{1}{x-1}Q\bigg]G \phantom{a} {\rm and} \phantom{a} \frac{\partial G}{\partial \bar{x}}=G\bigg[\frac{1}{\bar{x}}P^t+\frac{1}{\bar{x}-1}Q^t\bigg],
\end{equation}
where the matrices $P$ and $Q$ are given by
\begin{equation}
  P= -\frac{1}{N(k+N)} \begin{pmatrix} \frac{2(N^2-1)}{3} & N\\0 & -\frac{N^2+2}{3} \end{pmatrix} \phantom{a} {\rm and} \phantom{a} Q= -\frac{1}{N(k+N)} \begin{pmatrix} -\frac{N^2+2}{3} & 0\\N & \frac{2(N^2-1)}{3}\end{pmatrix}.
\end{equation}
The general solution to these equations takes the form
\begin{equation}
   G_{AB}(x, \bar{x}) =  X_{ij} F^{i}_{A} (x) F^{j}_B (\bar{x} ),
 \end{equation}
 where the indices $i,j$ run over the primaries in the intermediate channel. These are  the identity $(\mathds{1} )$ and the adjoint $(\theta)$ fields. 
$F^{i}_{A}(x)$ are the conformal blocks 
\begin{eqnarray}
\nonumber
 F^{\mathds{1}}_1(x) &=& x^{-\frac{4h_g}{3}}(1-x)^{h_{{\theta}}-\frac{4h_g}{3}} F \left(\frac{1}{\tilde{k}},-\frac{1}{\tilde{k}};1-\frac{N}{\tilde{k}};x \right) \ , \\
 \nonumber
 F^{\mathds{\mathds{1}}}_2 (x)&=&  {1 \over k} x^{1-\frac{4h_g}{3}}(1-x)^{h_{{\theta}}-\frac{4h_g}{3}} F\left(1+\frac{1}{\tilde{k}},1-\frac{1}{\tilde{k}};2-\frac{N}{\tilde{k}};x \right) \ , \\
 \nonumber
  F^{\mathds{\theta}}_1(x)&=& x^{h_{{\theta}}-\frac{4h_g}{3}}(1-x)^{h_{{\theta}}-\frac{4h_g}{3}} F\left(\frac{N}{\tilde{k}}-\frac{1}{\tilde{k}},\frac{N}{\tilde{k}}+\frac{1}{\tilde{k}};1+\frac{N}{\tilde{k}};x \right) \ , \\ 
F^{\mathds{\theta}}_2 (x) &=& - Nx^{h_{\theta}-\frac{4h_g}{3}}(1-x)^{h_{\theta}-\frac{4h_g}{3}} F \left(\frac{N}{\tilde{k}}-\frac{1}{\tilde{k}},\frac{N}{\tilde{k}}+\frac{1} {\tilde{k}};\frac{N}{\tilde{k}};x \right) \ ,
\end{eqnarray}  
where $\tilde{k}=k+N$ and $F(a, b, c; x)$ is the Gauss hypergeometric function\footnote{Our conventions for the definition of the Gauss hypergeometric function will be same as that of ~\cite{Weisstein2}.}. We define the holomorphic and the anti-holomorphic blocks:
\begin{eqnarray}
\mathcal{F}^{\mathds{1}}(x) &=& I_1 F^{\mathds{1}}_1(x)  + I_2 F^{\mathds{1}}_2(x)\\
\bar{\mathcal{F}}^{\mathds{1}}(\bar{x}) &=& \bar{I}_1 F^{\mathds{1}}_1(\bar{x})  + \bar{I}_2 F^{\mathds{1}}_2(\bar{x}) \\
\mathcal{F}^{\theta} (x) &=&  I_1 F^{\mathds{\theta}}_1(x)  + I_2 F^{\theta}_2(x) \\
\bar{\mathcal{F}}^{\theta}(\bar{x}) &=& \bar{I}_1 F^{\theta}_1(\bar{x})  + \bar{I}_2 F^{\theta}_2(\bar{x}).
\end{eqnarray}
With this, the correlator factorises into holomorphic and anti-holomorphic parts:
\begin{equation}
\label{holfac}
    G^{\beta_1 \alpha_2 \alpha_3 \beta_4}_{\alpha_1 \beta_2 \beta_3 \alpha_4}(x,\bar{x}) = X_{ij} \mathcal{F}^{i}(x) \bar{\mathcal{F}}^{j}(\bar{x}).
\end{equation}

     As discussed in section \ref{review}, general correlators transform as a six dimensional modular vector under the action of the modular group. Just as in the correlator described above, there are two holomorphic and two anti-holomorphic blocks associated with each correlator. This implies that the vector valued modular form requires 24 coefficients for its specification. This number is large even if one wants to carry out modular averaging as per \pref{aver} numerically. Luckily, one can
simplify the computation by exploiting the fact that \pref{holfac} implies that the $X_{ij}$ are independent of the $SU(N)$ left and right tensor indices.  We 
will make choices for these so that the correlator has two pairs of identical operators i.e. we will take $\alpha_1=\alpha_4$, $\beta_1=\beta_4$, $\alpha_2=\alpha_3$, $\beta_2=\beta_3$. With this we have
\begin{equation}
\label{idef}
   I_1 = I_2 \equiv I \phantom{abc} \text{and}  \phantom{abc}  \bar{I}_1 = \bar{I}_2 \equiv \bar{I}.
\end{equation}
As a result, the six dimensional vector space collapses to a three dimensional one (after use of equation \pref{triv}):
\begin{equation}
\label{modvec}
\vec{G}  = \left( G^{\beta_1 \alpha_2 \alpha_2 \beta_1}_{\alpha_1 \beta_2 \beta_2 \alpha_1}(\tau,\bar{\tau}),   G^{\beta_1 \alpha_2 \alpha_1 \beta_2}_{\alpha_1 \beta_2 \beta_1 \alpha_2}(\tau,\bar{\tau}), G^{\beta_1 \alpha_1 \alpha_2 \beta_2}_{\alpha_1 \beta_1 \beta_2 \alpha_2}(\tau,\bar{\tau}) \right), 
  \end{equation}
its transformations under the modular group as given by \pref{master} reduces to
\begin{equation}
\label{colG}
  \begin{aligned}
    &\vec{G}(T\cdot \tau,T\cdot \bar{\tau})=\sigma(T)\cdot\vec{G}(\tau,\bar{\tau}),\\
    &\vec{G}(S\cdot \tau,S\cdot \bar{\tau})=\sigma(S)\cdot\vec{G}(\tau,\bar{\tau}),
  \end{aligned}
\end{equation}
where
\begin{equation}
\label{stm}
  \sigma(T)=\begin{pmatrix} 0 & 1 & 0\\1 & 0 & 0\\0 & 0 & 1 \end{pmatrix},\text{ }\sigma(S)=\begin{pmatrix} 1 & 0 & 0\\0 & 0 & 1\\0 & 1 & 0 \end{pmatrix}.
\end{equation}
We list the conformal blocks associated with the three correlators in \pref{modvec} and their transformation properties under the modular group
in Appendix \ref{apblocks}.

 We will  primarily perform the  modular averaging as per the algorithm in \pref{newav} (although also briefly consider averaging as per the prescription in \pref{aver} in
Appendix \ref{apmodav}). For the representation of
$PSL(2, \mathbb{Z})$ generated by the matrices in \pref{stm}, it is easy to see that the vector $(1, 0, 0)$ is left
invariant by the subgroup generated by the actions of $S$ and $T^{2}$. This is called the theta group \cite{Bruggeman}. 
This subgroup is an index $3$ subgroup of $PSL(2,\mathbb{Z})$ which contains $\Gamma(2)$ as an index $2$
normal subgroup. In order to carry out the modular averaging as per \pref{newav}, we require the actions of the elements of this subgroup
on the conformal blocks associated with the stripped correlator $G^{\beta_1 \alpha_2 \alpha_2 \beta_1}_{\alpha_1 \beta_2 \beta_2 \alpha_1}(\tau,\bar{\tau})$. These blocks are 
 \begin{eqnarray}
 \label{newblock}
 \nonumber
\mathcal{H}^{\mathds{1}}(x) &=& I F^{\mathds{1}}_1(x)  + I F^{\mathds{1}}_2(x) \\
 \mathcal{H}^{\theta} (x) &=&  I F^{\mathds{\theta}}_1(x)  + I F^{\theta}_2(x),
  \end{eqnarray}
 with $I$ and $\bar{I}$ as defined in \pref{idef}.

The transformation properties of these blocks under $S$ and $T^{2}$  can be obtained  from Appendix \ref{apblocks}.  The action of $T^2$ is given by
 \begin{equation}
\mathcal{H}_{i} \left(T^2.x \right) = \mathcal{H}_{j} \left(x \right) M_{ji}(T^2),
 \end{equation}
 where
\begin{equation}
\label{tmat}
  M(T^2)= e^{-i 4\pi(N^2-1)/3N\tilde{k}}
  \begin{pmatrix}
    1 & 0\\
    0 & e^{i 2\pi N/\tilde{k}}
  \end{pmatrix}.
\end{equation}  
The action of $S$ is given by
 \begin{equation}
\mathcal{H}_{i} \left(S.x \right) = \mathcal{H}_{j} \left(x \right) M_{ji}(S),
 \end{equation}
 where
 \begin{equation}
\label{smat}
  M(S)=
  \begin{pmatrix}
    -\frac{\tilde{k}\Gamma \left({N} {/}{\tilde{k}} \right)\Gamma \left({k}{/}{\tilde{k}}\right)}{\Gamma \left({1}{/}{\tilde{k}}\right)\Gamma \left(-{1}{/}{\tilde{k}}\right)} & -\frac{N\Gamma^2 \left({N}{/}{\tilde{k}}\right)}{\Gamma \left({N}{/}{\tilde{k}}-{1}{/}{\tilde{k}}\right)\Gamma \left({N}{/}{\tilde{k}}+{1}{/}{\tilde{k}}\right)}\\
    -\frac{\Gamma^2 \left({k}{/}{\tilde{k}}\right)}{N\Gamma \left(k/ \tilde{k}  - 1/ \tilde{k} \right)\Gamma \left(k/ \tilde{k}  + 1/ \tilde{k} \right)} & \frac{\tilde{k}\Gamma \left({N}{/}{\tilde{k}}\right)\Gamma \left({k}{/}{\tilde{k}}\right)}{\Gamma \left({1}{/}{\tilde{k}}\right)\Gamma \left(-{1}{/}{\tilde{k}}\right)}
  \end{pmatrix}.
\end{equation}
 Successive actions of $M(T^2)$  and $M(S)$ can be used to obtain the action of any element $\gamma$ of the theta subgroup of the modular group on $\mathcal{H}_{i}(x)$,
 we shall denote the associated matrix by $M(\gamma)$. With the definitions in \pref{newblock}, the most general form of solutions to the KZ equations with two identical operators can be written as
\begin{equation}
  G^{\beta_1 \alpha_2 \alpha_2 \beta_1}_{\alpha_1 \beta_2 \beta_2 \alpha_1}(x,\bar{x}) = X_{ij} \mathcal{H}^{i}(x) \bar{\mathcal{H}}^{j}(\bar{x}).
\end{equation}
Under the action of an element $\gamma$ of the theta subgroup, the matrix $X$ transforms as
\begin{equation}
\label{xtran} 
X \to M(\gamma) X M^{\dagger}(\gamma).
\end{equation}
We note that under composition
\begin{equation}
\label{compo}
 M(\gamma_2 .\gamma_1) = M(\gamma_1). M(\gamma_2).
\end{equation}
\section{Correlators from Modular Averaging}
\label{Saverage}

   Having obtained the transformation properties of the conformal blocks we now turn to constructing correlators from modular averaging. In this section, we will carry out the 
modular averaging  as per the prescription in \pref{newav}. As described in the previous section, we will focus on the correlator \pref{maincor} after making choices for $SU(N)$ left and right indices so that two pairs of operators are identical.
$G^{\rm light}$ will be taken to be the contribution of the vacuum conformal block, as in  \cite{Maloney:2017} we will refer to this as the 
seed contribution. The transformation \pref{xtran} of the matrix $X$ implies that one can write the result of modular averaging as
\begin{equation}
\label{ourav}
  X^{\rm av}=\mathcal{N}^{-1} \cdot \sum_{\gamma\in\text{ }\Gamma} M(\gamma) \cdot C_{\rm seed} \cdot M(\gamma)^\dagger,
\end{equation}
where we have used $\Gamma$ to denote the theta subgroup and
\begin{equation}
 C_{\rm seed}=\begin{pmatrix} 1 & 0 \\ 0 & 0 \end{pmatrix}.
\end{equation}
The  normalization constant $\mathcal{N}$ is determined by demanding $[X]_{11}=1$, so that the $x \to 0$ behaviour of the  correlator is
correct. For comparison we record the (exact)result of \cite{KZ:1984}:
\begin{equation}
\label{kzr}
  X^{\rm KZ} =\begin{pmatrix} 1 & 0 \\ 0 & \frac{\Gamma \left({N}{/}{\tilde{k}}-{1}{/}{\tilde{k}}\right)\Gamma \left({N}{/}{\tilde{k}}+{1}{/}{\tilde{k}}\right){\Gamma^2 \left(1-{N}{/}{\tilde{k}}\right)}}{N^2\Gamma \left(1-{N}{/}{\tilde{k}}+{1}{/}{\tilde{k}}\right)\Gamma \left(1-{N}{/}{\tilde{k}}-{1}{/}{\tilde{k}}\right){\Gamma^2 \left({N}{/}{\tilde{k}}\right)}} \end{pmatrix}.
  \end{equation}

  Before carrying out the sum in explicit examples, let us discuss some generalities.  Any element of $\Gamma$ can be expressed as
\begin{equation}
      \gamma=T^{2n_1}ST^{2n_2}S \cdots ST^{2n_k},
\end{equation}  
for some choice of integers $n_i$ (see e.g. \cite{Francesco}).   Since we are dealing with a normalised sum, the sum can be
reduced to be over the orbit of $C_{\rm seed}$. Given this, our interest shall be in $\gamma$ whose action will generate
distinct elements. In this context, note that for all  $(N,k)$ the action of
$M(T^2)$ on $C_{\rm seed}$ is trivial. Also, in the representations under consideration (which are given in \pref{tmat}), $T^2$ has finite order. Thus,
all distinct $M(\gamma)$ can be generated by considering non-negative values of $n_{i}$ upto the order of $T^2$. Furthermore, for $M(\gamma)$ of the form $e^{i \alpha} \mathds{1}$, its  action  \pref{xtran} on any X is trivial. We define $m(N,k)$ as the smallest positive integer such that
\begin{equation}
\label{mdef}
M(T^{2m(N,k)}) \propto \mathds{1}.
\end{equation}
 With this, given the trivial  actions described above, a list of  $\gamma$s whose actions contain the orbit of
 $C_{\rm seed}$  can be constructed by considering all elements of the form
\begin{equation}
\label{list}
  \gamma=S T^{2r_1}S  \cdots S T^{2r_\ell},
\end{equation}
 with $\ell$ taking values over natural numbers, $r_{i} = 1 \cdots (m-1)$ for $i = 1 \cdots (\ell -1)$ and
 $r_{\ell} = 0 \cdots (m-1)$.   We define the length of an element in the list to be
 the  value of $\ell$ associated with it (and denote it as $\ell(\gamma)$). The composition rule \pref{compo} implies
\begin{equation}
\label{mlist}
  M(\gamma)= M(T^{2r_\ell})M(S)  \cdots M(S)M(T^{2r_1})M(S).
\end{equation}

  If the stabilser of $C_{\rm seed}$ under the action $C_{\rm seed} \to M(\gamma) \cdot C_{\rm seed} \cdot M(\gamma)^\dagger$ has finite index, then the 
sum reduces to a finite number of terms. Otherwise, one has to deal with an infinite sum. We begin by discussing some models in which the stabiliser is of finite index.

   Models with $N=k$ are particularly simple. For $N=k$, the actions of $S$ and $T$  as given by \pref{smat} and \pref{tmat} can be written as
\begin{equation}
M(S)= \begin{pmatrix} 
\sin\frac{\pi}{2k} & -k\cos\frac{\pi}{2k}\\ -\frac{1}{k}\cos\frac{\pi}{2k} & -\sin\frac{\pi}{2k} 
\end{pmatrix} \ , 
\phantom{abcd}
M(T^2)= {e}^{  {-\frac{2\pi i}{3}. \frac{(N^2 -1)}{N^2}}  }\begin{pmatrix} 1& 0\\ 0 & -1
\end{pmatrix}.
\end{equation}
Note that $M(T^{4}) \propto \mathds{1}$, thus the highest power of $T$ that needs to be included while generating
the matrices $M(\gamma)$ in the list in \pref{list} is $T^2$.  Let us start by discussing a particular example.

\vspace{0.6 cm}

\noindent \underline{$N=3, k=3 $} :  For $N=3, k=3$,  the matrices $M(S)$ and $M(T^2)$ are 
\begin{equation}
M(S)  =  \left(
\begin{array}{cc}
 \frac{1}{2} & -\frac{3 \sqrt{3}}{2}  \\
 -\frac{1}{2 \sqrt{3}} & -\frac{1}{2} \\
\end{array}
\right),
\phantom{abcd}
M(T^2)  = {e}^{  {-\frac{16\pi i}{27}} }\left(
\begin{array}{cc}
 1 & 0  \\
 0 & -1 \\
\end{array}
\right)
\end{equation}
The orbit of $C_{\rm seed}$ consists of three matrices. It is generated by the action of $\mathds{1}, S$ and $ST^2$. We tabulate the results of these actions
in  Table~\ref{tab:1}. The normalised sum over the orbit \pref{ourav} reproduces the KZ result.

\begin{table}[H]
\begin{center}
  \begin{tabular}{|c|c|}
    \hline
    $\gamma$ & $\text{ } M(\gamma) \cdot C_{\rm seed} \cdot M(\gamma)^{\dagger} \text{ }$\\
    \hline
    $\mathds{1}$ & $\begin{pmatrix} 1 & 0 \\ 0 & 0 \end{pmatrix}$\\[0.4cm]
    $S$ & $\begin{pmatrix} \frac{1}{4} & -\frac{1}{4\sqrt{3}} \\ -\frac{1}{4\sqrt{3}} & \frac{1}{12} \end{pmatrix}$\\[0.4cm]
    $ST^2$ & $\begin{pmatrix} \frac{1}{4} & \frac{1}{4\sqrt{3}} \\ \frac{1}{4\sqrt{3}} & \frac{1}{12} \end{pmatrix}$\\[0.4cm]
    \hline
    $X^{\rm av}$  & $\begin{pmatrix} 1 & 0 \\ 0 & \frac{1}{9} \end{pmatrix}$\\[0.4cm]
    \hline
  \end{tabular}
\end{center}
\caption{ Orbit of the vacuum block  for  $N=3, k=3$}
\label{tab:1}
\end{table}

  For general values  $N$ ($=k$), one can show that the orbit of $C_{\rm seed}$ is finite by taking repeated products of the matrices $M(S)$ and $M(T^2)$. The orbit is the set
\begin{equation}
\label{orbneqk}
  \bigg\{ \begin{pmatrix} \sin^2 \alpha & -\frac{1}{k}\sin\alpha\cos\alpha \\ -\frac{1}{k}\sin\alpha\cos\alpha & \frac{1}{k^2}\cos^2\alpha \end{pmatrix} \bigg\}
\end{equation}
where $ \alpha = {\pi (2s +1)\over 2 k} $ with $s = 0 \cdots (k-1)$ for $k$ odd, and  $ \alpha = {\pi s\over 2k}$ with $s = 0 \cdots (2k-1)$ for $k$ even (we derive  this in Appendix \ref{neqk}).

The sums over the orbits can be performed using the identities
$$
\sum_{s=0}^{k-1} \sin^2{\pi (2s +1)\over 2 k}={k\over 2}=\sum_{s=0}^{k-1} \cos^2{\pi (2s +1)\over 2 k}, \quad \sum_{s=0}^{k-1} \sin{\pi (2s +1)\over k}=0
$$
for $k$ odd and 
$$\sum_{s=0}^{2k-1} \sin^2{\pi s\over 2 k}=k=\sum_{s=0}^{2k-1} \cos^2{\pi s\over 2 k}, \quad \sum_{s=0}^{2k-1} \sin{\pi s\over k}=0$$
for $k$ even. Normalising the sum, one finds
\begin{equation}
 X^{\rm av} = 
 \begin{pmatrix}
 1 & 0 \\
 0 & 1/ k^2
 \end{pmatrix},
\end{equation}
which is in agreement with \pref{kzr}.

   We now turn to models with $N \neq k$ models with finite orbits. For $k=1$ and any finite $N$ the  actions of $S$ and $T^2$  as given by \pref{smat} and \pref{tmat} 
take the identity block to a multiple of itself. Thus the adjoint block decouples and upon modular averaging the correlator is given by  $|\mathcal{F}_{\mathds{1}}^1(\tau)|^2$,
in keeping with \cite{KZ:1984}.  Next, we discuss two models: $N = 4, k = 2$ and $N = 2, k =4$. These examples will reappear in our discussion of the properties of modular averaging under interchange of $N$ and $k$ in section \ref{Snk}.

\noindent \underline{$N=4, k=2 $:} For $N=4, k=2$ we note that $M(T^6)\propto \mathds{1}$. The orbit of $C_{\rm seed}$ consists of four matrices. It is generated by the action of $\mathds{1}$, $S$, $ST^2$ and $ST^4$. The normalised sum over the orbit \pref{ourav} reproduces the KZ result which is $\frac{1}{16 \sqrt[3]{2}}$.

\noindent \underline{$N=2, k=4 $:} For $N=2, k=4$ we note that $M(T^6)\propto \mathds{1}$. The orbit of $C_{\rm seed}$ consists of four matrices. It is generated by the action of $\mathds{1}$, $S$, $ST^2$ and $ST^4$. The normalised sum over the orbit \pref{ourav} reproduces the KZ result which is $\frac{1}{2 \sqrt[3]{4}}$.

  Finally, we present some models whose orbits do not seem to be finite. We will analyse the models numerically. As described
in our general discussion in the beginning of the section, a list of $\gamma$s whose actions contain the orbit of
$C_{\rm seed}$ can be obtained by considering elements of the form \pref{list}. To implement the numerics, we will
organise the sum over the actions of the elements of the list in terms of the length of the elements. We define\footnote{Our implementation of the numerics is similar to \cite{Maloney:2017}.}
\begin{equation}
\label{numereq}
  X^{\rm{av}}(\ell_{\rm max})=\mathcal{N}(\ell_{\rm max})^{-1} \cdot \sum'_{\ell(\gamma) \leq \ell_{\rm max}} M(\gamma) \cdot C_{\rm seed} \cdot M(\gamma)^\dagger ,
\end{equation}
where the primed sum indicates that we include distinct elements of the orbit of $C_{\rm seed}$  in the sum. The normalisation constant $\mathcal{N}(\ell_{\rm max})$ is determined
by requiring $X_{11}^{\rm{av}}(\ell_{\rm max}) =1$, so that the $x \to 0$ behaviour of the correlator is correctly reproduced at every value of $\ell_{\rm max}$.

\vspace{0.6cm}

\begin{figure}[H]
  \centering
    \includegraphics[width=8.0cm]{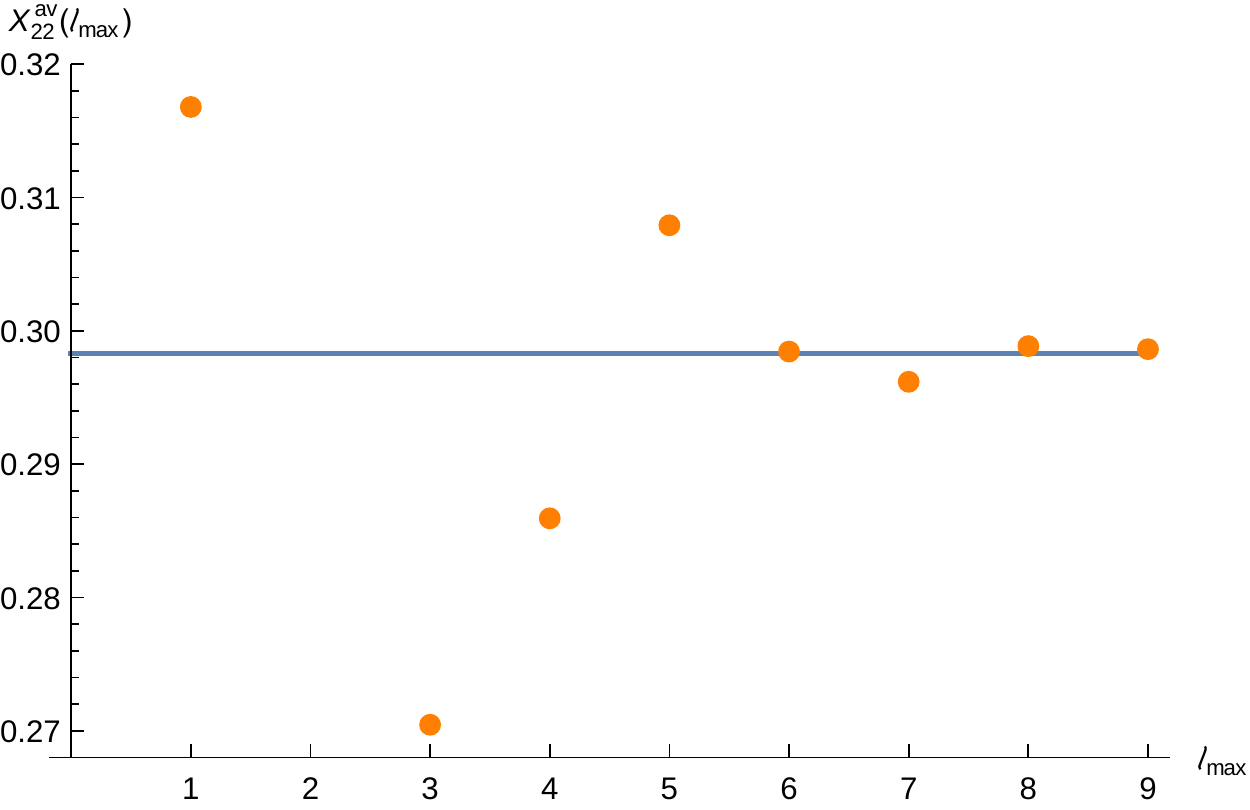}
  \caption{Orange dots show $X^{\rm av}_{22}(\ell_{\rm max})$ in the range $[0.268, 0.320]$ plotted against $\ell_{\rm max}$. Blue horizontal line at $0.29831$ represents $X^{\rm KZ}_{22}$.}
\label{fig:1}
\end{figure}

\noindent \underline{$N=2$, $k=3$:} For $N=2, k =3$, we have performed sum in \pref{numereq} upto $\ell_{\rm max}=9$. This
involves $429226$ distinct contributions to the sum. We find 
$X^{\rm av}_{22}(9) = 0.29863$, which is in good agreement with the exact result  \pref{kzr}, $X^{\rm KZ}_{22} \approx 0.29831$.
The off diagonal entries of $X^{\rm av}(9)$ are of the order of $10^{-13}$. Figure ~\ref{fig:1} shows our results for $X^{\rm av}_{22}(\ell_{\rm max})$ as a function of $\ell_{\rm max}$.
Note that  $X^{\rm av}_{22}(\ell_{\rm max})$ approaches the exact result in an oscillatory manner.  Prior to normalisation of the sum, both the $(1,1)$-element as well as the $(2,2)$-element of the matrix  have approximately linear growths (all terms in the sum make positive definite contributions to these elements).
However, as exhibited by the plot, the ratio of  the two quantities (which is $X^{\rm av}_{22}(\ell_{\rm max})$) tends to a constant. Off-diagonal entries are small
as a result of phase cancellations.

\begin{figure}[H]
  \centering
    \includegraphics[width=8.0cm]{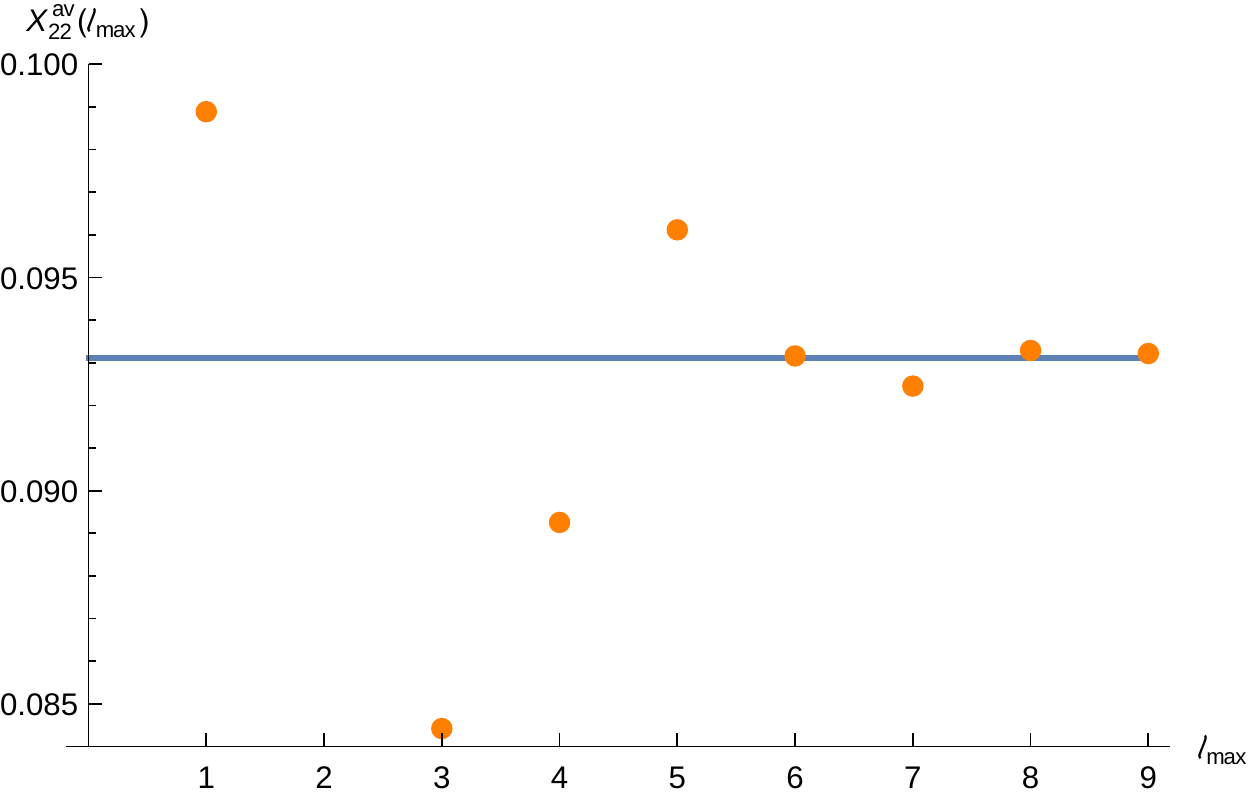}
  \caption{Orange dots show $X^{\rm av}_{22}(\ell_{\rm max})$ in the range $[0.084, 0.100]$ plotted against $\ell_{\rm max}$. Blue horizontal line at $0.0931172$ represents $X^{\rm KZ}_{22}$.}
\label{fig:2}
\end{figure}

\noindent \underline{ $N=3$, $k=2$:} For $N=3, k =2$, we have performed sum in \pref{numereq} upto $\ell_{\rm max}=9$. This
involves $429226$ distinct contributions to the sum. We find 
$X^{\rm av}_{22}(9) = 0.0932166$, which is in good agreement with the exact result  \pref{kzr}, $X^{\rm KZ}_{22} \approx 0.0931172$.
The off diagonal entries of $X^{\rm av}(9)$ are of the order of $10^{-14}$. Figure ~\ref{fig:2} shows our results for $X^{\rm av}_{22}(\ell_{\rm max})$ as a function of $\ell_{\rm max}$. As in the previous example, $X^{\rm av}_{22}(\ell_{\rm max})$ approaches the exact result in an oscillatory manner.  Other features of the numerics are also similar \footnote{This is also true for all models that we study numerically.}.

\begin{figure}[H]
  \centering
    \includegraphics[width=8.0cm]{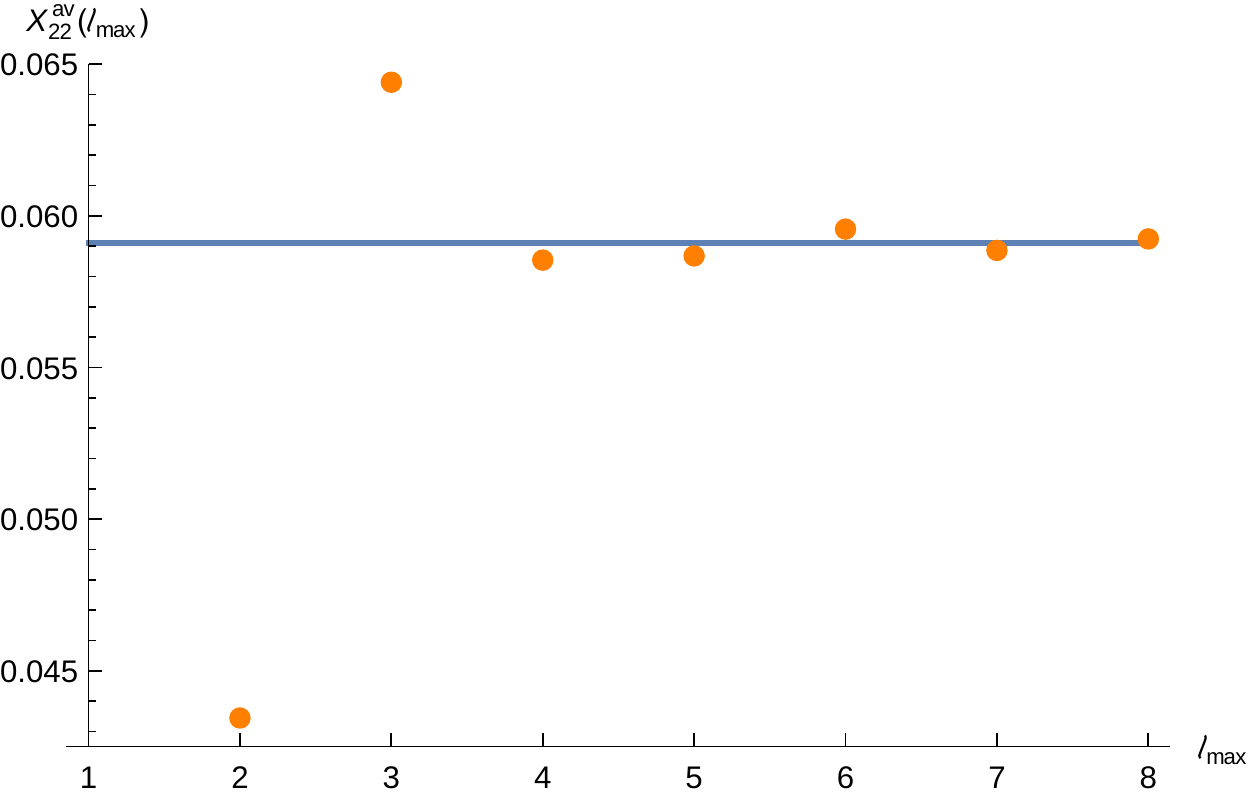}
  \caption{Orange dots show $X^{\rm av}_{22}(\ell_{\rm max})$ in the range $[0.0425, 0.0650]$ plotted against $\ell_{\rm max}$. Blue horizontal line at $0.0591147$ represents $X^{\rm KZ}_{22}$.}
\label{fig:3}
\end{figure}

\noindent \underline{ $N=4$, $k=3$:} For $N=4, k =3$, we have performed sum in \pref{numereq} upto $\ell_{\rm max}=8$. This
involves $2338785$ distinct contributions to the sum. We find 
$X^{\rm av}_{22}(8) = 0.0592407$, which is in good agreement with the exact result  \pref{kzr}, $X^{\rm KZ}_{22} \approx 0.0591147$.
The off diagonal entries of $X^{\rm av}(8)$ are of the order of $10^{-14}$. Figure ~\ref{fig:3} shows our results for $X^{\rm av}_{22}(\ell_{\rm max})$ as a function of $\ell_{\rm max}$.

\begin{figure}[H]
  \centering
    \includegraphics[width=8.0cm]{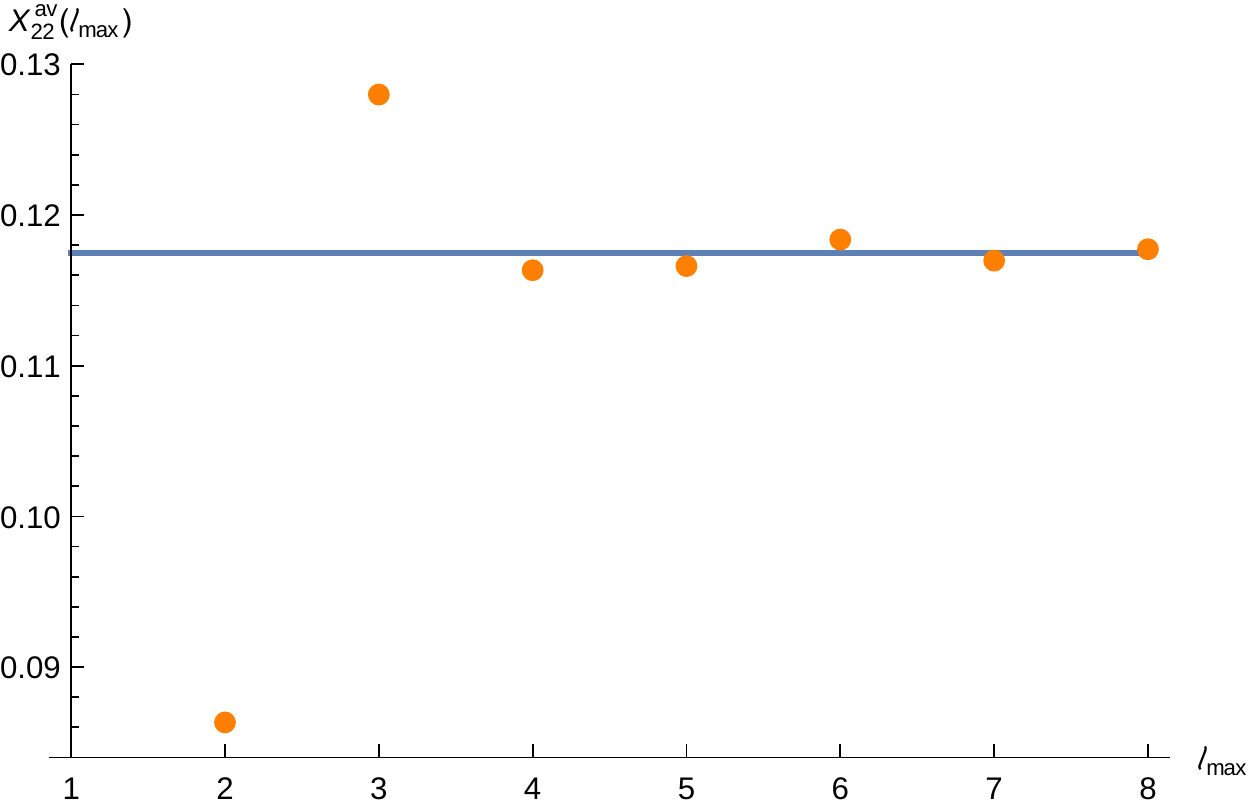}
  \caption{Orange dots show $X^{\rm av}_{22}(\ell_{\rm max})$ in the range $[0.084, 0.130]$ plotted against $\ell_{\rm max}$. Blue horizontal line at $0.117474$ represents $X^{\rm KZ}_{22}$.}
\label{fig:4}
\end{figure}

\noindent \underline{ $N=3$, $k=4$:} For $N=3, k =4$, we have performed sum in \pref{numereq} upto $\ell_{\rm max}=8$. This
involves $2338785$ distinct contributions to the sum. We find 
$X^{\rm av}_{22}(8) = 0.117725$, which is in good agreement with the exact result  \pref{kzr}, $X^{\rm KZ}_{22} \approx 0.117474$.
The off diagonal entries of $X^{\rm av}(8)$ are of the order of $10^{-14}$. Figure ~\ref{fig:4} shows our results for $X^{\rm av}_{22}(\ell_{\rm max})$ as a function of $\ell_{\rm max}$.

As the values of $N$ and $k$ are increased the numerics can become quite involved. Getting accurate results might require large values of $\ell_{\rm max}$. Models with $(N,k)$ equals to $(5,6)$ and $(6,5)$ provide examples of this. We discuss them in Appendix \ref{exnneqk}.

  Finally, we have also considered the prescription for  constructing correlators by averaging   over the whole $PSL(2, \mathbb{Z})$  \pref{aver}. This involves
averaging over a vector and  hence is more complicated. We briefly present our results on this in Appendix \ref{apmodav} and leave more detailed explorations
for the future.

   In summary, in all the cases that we have examined, modular averaging successfully reproduces the result of \cite{KZ:1984}. The correlators can be considered as extremal in the sense of \cite{Maloney:2017}. For extremal correlators,  modular averaging sums can be thought
of as providing an alternate prescription for their computation. Next, we will examine the properties of these sums involved under 
interchange of $N$ and $k$.

\section{$N \leftrightarrow k$ in Modular Averages}
\label{Snk}

   As described in the introduction, an interesting property of  WZW models is level rank duality. 
   In this section, we will show that there is a simple one to one correspondence between  individual terms in the modular averaging sums for  correlators in the $(N,k)$ and $(k,N)$ theories. 
   
     We will be simultaneously dealing with the $(N,k)$ and $(k,N)$ theories in this section, let us begin by introducing notation adapted for the purpose. We 
will include labels in the matrices \pref{tmat} and \pref{smat} which generate the actions of $S$ and $T^2$, to indicate the theory they belong to.
\begin{equation}
\label{tmatnk}
  M_{N,k}(T^2)= e^{-i 4\pi(N^2-1)/3N\tilde{k}}
  \begin{pmatrix}
    1 & 0\\
    0 & e^{i 2\pi N/\tilde{k}}
  \end{pmatrix}
   \equiv 
   e^{i \alpha(N,k)}  
     \begin{pmatrix}
    1 & 0\\
    0 & e^{i\phi(N,k)}
  \end{pmatrix}
  \end{equation}  
and 
 \begin{equation}
\label{smatnk}
  M_{N,k}(S)=
  \begin{pmatrix}
    -\frac{\tilde{k}\Gamma \left({N} {/}{\tilde{k}} \right)\Gamma \left({k}{/}{\tilde{k}}\right)}{\Gamma \left({1}{/}{\tilde{k}}\right)\Gamma \left(-{1}{/}{\tilde{k}}\right)} & -\frac{N\Gamma^2 \left({N}{/}{\tilde{k}}\right)}{\Gamma \left({N}{/}{\tilde{k}}-{1}{/}{\tilde{k}}\right)\Gamma \left({N}{/}{\tilde{k}}+{1}{/}{\tilde{k}}\right)}\\
    -\frac{\Gamma^2 \left({k}{/}{\tilde{k}}\right)}{N\Gamma \left(k/ \tilde{k}  - 1/ \tilde{k} \right)\Gamma \left(k/ \tilde{k}  + 1/ \tilde{k} \right)} & \frac{\tilde{k}\Gamma \left({N}{/}{\tilde{k}}\right)\Gamma \left({k}{/}{\tilde{k}}\right)}{\Gamma \left({1}{/}{\tilde{k}}\right)\Gamma \left(-{1}{/}{\tilde{k}}\right)}
  \end{pmatrix}
  \equiv
  \begin{pmatrix}
  a_s(N,k) & b_s(N,k) \\
  c_s(N,k) & d_s(N,k)
  \end{pmatrix}.
 \end{equation} 
We note that $d_{s}(N,k) = - a_s(N,k)$ and $b_{s}(N,k) . c_{s}(N,k) = 1 + a_s(N,k) . d_s(N,k)$. Also, $a_{s}(N,k)$ and the product $b_{s} (N,k) . c_s(N,k)$ are symmetric under the interchange of $N$ and $k$, i.e.
\begin{equation}
\label{srel}
  a_{s}(N,k) =  a_s(k,N),   \phantom{abc} d_{s}(N,k) = d_{s}(k,N), \phantom{abc} b_{s} (N,k) . c_s(N,k) = b_{s} (k,N) . c_s(k,N).
\end{equation}
Recall that the matrices given in \pref{mlist} provide a list whose actions contain the orbit of $C_{\rm seed}$. We will denote the matrices in the list by
\begin{equation}
\label{mmdef}
M_{N,k}^{\ell}(r_{1}, r_2 \cdots, r_\ell ) \equiv   M_{N,k}^{\ell}(r_{i}) \equiv M_{N,k}(T^{2r_\ell})M_{N,k}(S)  \cdots M_{N,k}(S)M_{N,k}(T^{2r_1})M_{N,k}(S).
\end{equation}
Note that with this $M^{\ell}_{N,k}(r_i)$ is a function of $r_1, r_2 \cdots r_{l}$; with $r_{i} = 1 \cdots \left(m(N,k)-1 \right)$ for $i = 1 \cdots (\ell -1)$ and
 $r_{\ell} = 0 \cdots \left(m(N,k)-1 \right)$ with $m(N,k)$ as defined in \pref{mdef}. We define $M^{0}_{N,k}$ to be the identity matrix.
 We now introduce another set of matrices 
\begin{equation}
\label{mmtdef}
  \tilde{M}^{\ell}_{N,k} (p_{1}, p_2 \cdots, p_\ell ) \equiv  \tilde{M}_{N,k}^{\ell}(p_{i}) \equiv M_{N,k}(T^{-2p_\ell})M_{N,k}(S)  \cdots M_{N,k}(S)M_{N,k}(T^{-2p_1})M_{N,k}(S).\end{equation}
$\tilde{M}^{\ell}_{N,k}(p_i)$ is a function of $p_1, p_2 \cdots p_{l}$; with $p_{i} = 1 \cdots \left(m(N,k)-1 \right)$ for $i = 1 \cdots (\ell -1)$ and
 $p_{\ell} = 0 \cdots \left(m(N,k)-1 \right)$. We will define $\tilde{M}^{0}_{N,k}$ to be the identity matrix.

    At any given length $\ell$, the set of matrices generated from the action of $M_{N,k}^{\ell}(r_{i})$ on $C_{\rm seed}$ is exactly same as the set generated from the action of  $\tilde{M}_{N,k}^{\ell}(p_{i})$ on $C_{\rm seed}$ i.e.
 \begin{eqnarray}
 \label{mtmequi}
 \nonumber
   \bigg\{ {M}_{N,k}^{\ell}(r_{i}) C_{\rm seed} {M}_{N,k}^{\dagger \ell}(r_{i});   r_{i} = 1 \cdots \left(m(N,k)-1 \right) \ \text{for} & \ i = 1 \cdots (\ell -1),
 r_{\ell} = 0 \cdots \left(m(N,k)-1 \right)\bigg\} \\
=   \bigg\{ \tilde{{M}}_{N,k}^{\ell}(p_{i}) C_{\rm seed} \tilde{{M}}_{N,k}^{\dagger \ell}(p_{i});  p_{i} = 1 \cdots \left(m(N,k)-1 \right) \ \text{for}& \ i = 1 \cdots (\ell -1), 
 p_{\ell} = 0 \cdots \left(m(N,k)-1 \right)) \bigg\}. \nonumber
 \\
 \end{eqnarray}
 This is a consequence of the fact that for any $X$ following equality (between sets) holds
 \begin{equation}
    \bigg\{ M_{N,k}( T^{2r}) X M^{\dagger}_{N,k}( T^{2r}); r = 0 \cdots  \left( m(N,k) - 1 \right)  \bigg\} =   \bigg\{ M_{N,k}( T^{-2p}) X M^{\dagger}_{N,k}( T^{-2p});  p = 0 \cdots  \left(m(N,k) -1\right) \bigg\}
 \end{equation}

     Given the equivalence in \pref{mtmequi}, while carrying out modular averaging, either set can be used to generate the sum over the orbit of $C_{\rm seed}$. While establishing the relationship between the modular averages in the $(N,k)$ and $(k,N)$ theories, it will be useful to generate the orbit for the $(N,k)$ theory  using the $M_{N,k}^{\ell}$ matrices and for the $(k,N)$ theory using $\tilde{M}_{k,N}^{\ell}$ matrices. The essential point will be to establish that the actions of the two matrices\footnote{Note since ${\rm{gcd}}(k+N,N) = {\rm{gcd}}(k,N) = {\rm{gcd}}(k+N, k)$,  $m(N,k) = m(k,N)$. This implies that the arguments of $M_{N,k}^{\ell}$ and $\tilde{M}_{k,N}^{\ell}$ take the same values.}
 \begin{equation}
     M_{N,k}^{\ell}(r_{1}, r_{2} \cdots r_{\ell})  \  \  \text{and} \ \  \tilde{M}_{k,N}^{\ell}(r_{1}, r_{2} \cdots r_{\ell})
 \end{equation}
 on $C_{\rm seed}$ are closely related. Let us begin by looking at the general from of the matrices $M_{N,k}^{\ell}(r_{1}, r_{2} \cdots r_{\ell})$ and 
 $\tilde{M}_{N,k}^{\ell}(r_{1}, r_{2} \cdots r_{\ell})$ . As shown in Appendix \ref{matap}, they can be written as 
\begin{equation}
\label{mpar}
   M_{N,k}^{\ell}(r_{1}, \cdots r_{\ell}) = {\rm exp} \left( i \alpha(N,k)  (\sum  r_i) \right) 
   \begin{pmatrix}
       {a}_{N,k}^{\ell}(r_1, \cdots  r_{\ell}) & b_s(N,k) {b}_{N,k}^{\ell}(r_1,  \cdots  r_{\ell}) \\
       c_s(N,k) {c}_{N,k}^{\ell}( r_1,  \cdots  r_{\ell}) & {d}^{\ell}_{N,k}( r_1,  \cdots r_{\ell}) 
    \end{pmatrix}
\end{equation}
\begin{equation}
\label{tmpar}
   \tilde{M}_{N,k}^{\ell}(r_{1}, \cdots r_{\ell}) = {\rm exp} \left(- i \alpha(N,k)  (\sum  r_i) \right) 
   \begin{pmatrix}
       \tilde{a}_{N,k}^{\ell}(r_1, \cdots  r_{\ell}) & b_s(N,k) \tilde{{b}}_{N,k}^{\ell}(r_1,  \cdots  r_{\ell}) \\
       {c}_s(N,k) \tilde{c}_{N,k}^{\ell}( r_1,  \cdots  r_{\ell}) & \tilde{d}^{\ell}_{N,k}( r_1,  \cdots r_{\ell}), 
    \end{pmatrix}
\end{equation}
with the functions appearing above obeying the relationships
\begin{eqnarray}
\nonumber
\tilde{a}_{k,N}^{\ell}(r_1, \cdots  r_{\ell}) &=& {a}_{N,k}^{\ell}(r_1, \cdots  r_{\ell}),  \phantom{abcd}
\tilde{b}_{k,N}^{\ell}(r_1, \cdots  r_{\ell}) = {b}_{N,k}^{\ell}(r_1, \cdots  r_{\ell}),  \\
\tilde{c}_{k,N}^{\ell}(r_1, \cdots  r_{\ell}) &=& {c}_{N,k}^{\ell}(r_1, \cdots  r_{\ell}),   \phantom{abcd}
\tilde{d}_{k,N}^{\ell}(r_1, \cdots  r_{\ell}) = {d}_{N,k}^{\ell}(r_1, \cdots  r_{\ell}).  
\label{exrelm}
\end{eqnarray}

   Now, let us discuss the implications of these relations for modular averages. As mentioned before, we 
will generate the orbit of the $(N,k)$ theory using the matrices $M_{N,k}^{\ell}$ and the $(k,N)$ theory
using the $\tilde{M}_{k,N}^{\ell}$ matrices. Firstly, note that \pref{mpar} and \pref{tmpar} imply that any duplications in the action of 
$M_{N,k}^{\ell}$ on $C_{\rm seed}$ implies a duplication in the action of $\tilde{M}_{k,N}^{\ell}$ on $C_{\rm seed}$
and vice versa\footnote{This together with \pref{mtmequi} explains  why the number of duplicates for theories related under $N \leftrightarrow k$ were same in our numerical analysis in section \ref{Saverage}.} i.e.
\begin{equation}
\label{fdupli}
  M^{\ell}_{N,k} (r_{i} ) C_{\rm seed} M^{\dagger \ell}_{N,k} (r_{i}) =  M^{\ell}_{N,k} (s_{i} ) C_{\rm seed} M^{\dagger \ell}_{N,k} (s_{i})
  \Longleftrightarrow
  \tilde{M}^{\ell}_{k,N} (r_{i} ) C_{\rm seed} \tilde{M}^{\dagger \ell}_{k,N} (r_{i}) =  \tilde{M}^{\ell}_{k,N} (s_{i} ) C_{\rm seed} \tilde{M}^{\dagger \ell}_{k,N} (s_{i})
\end{equation}
Furthermore,  we have
\begin{equation}
\label{cornorm}
   M^{\ell}_{N,k} (r_{i} ) C_{\rm seed} M^{\dagger \ell}_{N,k} (r_{i}) \big{|}_{11} =  \tilde{M}^{\ell}_{k,N} (r_{i} ) C_{\rm seed} \tilde{M}^{\dagger \ell}_{k,N} (r_{i}) \big{|}_{11}
\end{equation}
and 
\begin{equation}
\label{corsum}
   c^{2}_s(k,N) M^{\ell}_{N,k} (r_{i} ) C_{\rm seed} M^{\dagger \ell}_{N,k} (r_{i}) \big{|}_{22}=
    c^{2}_s(N,k) \tilde{M}^{\ell}_{k,N} (r_{i} ) C_{\rm seed} \tilde{M}^{\dagger \ell}_{k,N} (r_{i}) \big{|}_{22}.
 \end{equation}
With this\footnote{It is easy to check that these relationships hold for the (4,2) and (2,4) models (which have finite orbits). For other models we have checked
them numerically.}, it is natural to pair the matrix
$$
M^{\ell}_{N,k} (r_{i} ) C_{\rm seed} M^{\dagger \ell}_{N,k} (r_{i}) 
$$
in the orbit of $C_{\rm seed}$ of the $(N,k)$ theory with the matrix
$$
\tilde{M}^{\ell}_{k,N} (r_{i} ) C_{\rm seed} \tilde{M}^{\dagger \ell}_{k,N} (r_{i}) 
$$
in the orbit of  $C_{\rm seed}$ of the $(k,N)$ theory. This establishes our one to one correspondence between the terms that appear in the modular averaging sums of the two theories. Note that \pref{cornorm} implies that the normalisations of both the sums are equal. With this, \pref{corsum} implies that the all paired terms in the sums contribute to the sums with the ratio
\begin{equation}
\label{ratfin}
{ c^{2}_s(N,k) \over c^{2}_s(k,N) }.
\end{equation}

   Of course, since the ratio is same for all the pairs, from the point of view of modular averaging one can trivially write the relation (even without performing the sums)
\begin{equation}
\label{ope}
 {  X_{\rm av}(N,k) \big{|}_{22} \over   X_{\rm av}(k,N) \big{|}_{22} } = { c^{2}_s(N,k) \over c^{2}_s(k,N) } = \frac{ k^2 \Gamma^4\left(k/\tilde{k}\right) \Gamma^2\left(N/\tilde{k}-1/\tilde{k}\right) \Gamma^2\left(N/\tilde{k}+1/\tilde{k}\right)}{N^2 \Gamma^2\left(k/\tilde{k}-1/\tilde{k}\right) \Gamma^2\left(k/\tilde{k}+1/\tilde{k}\right) \Gamma^4\left(N/\tilde{k}\right)}.
 \end{equation}
One can check by making use of gamma function identities that this is indeed consistent with the KZ result \pref{kzr}. Thus, the one to one correspondence between the terms in the two sums has given us
relations between  OPE coefficients in the theories (as OPE coefficients can be obtained by taking the small cross ratio limit of the expressions of the correlators in terms of  conformal blocks).

  It is natural to ask if the one to one correspondence between the terms in the modular averaging  sums  in the two theories has any physical interpretation. In this context, we note that  it was argued
    in \cite{Maloney:2017} that for ``heavy operators" the modular averaging for genus zero correlators can be interpreted
    as a semiclassical  $AdS_3$ dual computation. More specifically, if the operator dimensions are of the order of the central charge (c) of the  theory but
    less than $c/12$ then the bulk path integral  has saddles corresponding to geodesic  propagation of heavy particles between the operator insertion points in the
    boundary  \cite{Jackson:2014nla, Fitzpatrick:2015zha, Fitzpatrick:2015dlt, Hijano:2015qja, Hijano:2015rla, Chang:2015qfa, daCunha:2016crm, Balasubramanian:2017fan, Cresswell:2018mpj, Bhatta:2016hpz }.  Performing the sum over the saddles  incorporating the back reaction of the heavy particle geodesics on the geometry and exchange of light primaries, yields the sum over modular channels. But, the operators considered in this article cannot be made heavy in the semiclassical limit, since $h_{g}/ c \sim 1 / Nk$.  One possibility is that the situation is similar to \cite{Castro:2012}   where the  topological sectors for the saddle point sum was as given in the semi classical limit
 even in the quantum regime. In any case, a computation similar to ours for operators satisfying the heavy operator criterion
    should help reveal how level rank duality works from  a holographic point of view.

\section{Conclusions}
\label{Sconclusions}

  In this article, we have analysed correlators involving two fundamentals and two anti-fundamentals in $SU(N)$ WZW theories  using 
modular averaging. After determining  the transformations of the conformal blocks under  $S$ and $T$ transformations, correlators were expressed
as sum of the action of the elements of the theta subgroup of $PSL(2, \mathbb{Z})$ on the vacuum block. We found that for all models with $N=k$ the orbit of the vacuum block is finite and modular averaging reproduces the correlator correctly. In models where we were unable to characterise the orbit we performed the sums numerically; modular averaging  successfully reproduced the correlators, providing strong evidence that the correlators examined in this paper are extremal in the sense of \cite{Maloney:2017}. An important direction for future study is developing a better understanding of the modular averaging sums. This would
involve finding the  criterion which makes the orbit of the vacuum block finite and study  of convergence properties  when the orbit is not finite. 

   We have found a close relationship between modular averaging for correlators involving fundamentals and anti-fundamentals in the $(N,k)$ and $(k,N)$ theories. In section \ref{Snk}, we established a one two one correspondence
between the orbits of the vacuum conformal blocks of the two theories. The contributions of the paired terms to their respective sums was given by a ratio of elements
of  braids matrices in the theories.  This allowed us to obtain a simple relationship between OPE coefficients. A prescription relating  general correlators  of WZW models under level rank  duality  has  been given in    \cite{Schnitzer1}. The braid matrices of the theories for general correlators have been related in \cite{Schnitzer2, Naculich:1990}. It will be interesting to study the implications of these relations for modular averaging in more general correlators.

   As discussed in the later part of the previous section, we believe that our results give a strong hint that holographic computations can make various aspects of level rank duality in
WZW models manifest. A first step in this direction can be to consider  correlators of heavy operators in the theories and analyse their conformal
blocks in the semi-classical limit.

\section*{Acknowledgements}

We would like  to  thank Anirban Basu,  Dileep Jatkar, Henry Maxfield, Shiraz Minwalla, Satchi Naik,  Gim Seng Ng, Mahasweta Pandit and Ashoke Sen for discussions.

\appendix
\label{sec:9}

\section{Conformal Blocks and Their Transformations:}
\label{apblocks}

   In this Appendix, we list the conformal blocks associate with the following three correlators\footnote{The  other three independent correlators in \pref{gvec} are related to
these by the interchange $I_1 \leftrightarrow I_2$. Thus they can be easily obtained from the data in this Appendix.}
\bea
\label{corlist}
\langle  {g_{\alpha_1}}^{\beta_1}(z_1,\bar{z}_1) \cdot {g^{-1}_{\beta_2}}^{\alpha_2}(z_2,\bar{z}_2) \cdot {g^{-1}_{\beta_3}}^{\alpha_3}(z_3,\bar{z}_3) \cdot {g_{\alpha_4}}^{\beta_4}(z_4,\bar{z}_4) \rangle  \\
\langle  {g_{\alpha_1}}^{\beta_1}(z_1,\bar{z}_1) \cdot {g^{-1}_{\beta_2}}^{\alpha_2}(z_2,\bar{z}_2) \cdot {g_{\alpha_4}}^{\beta_4}(z_3,\bar{z}_3) \cdot {g^{-1}_{\beta_3}}^{\alpha_3}(z_4,\bar{z}_4) \rangle \\
\langle  {g_{\alpha_1}}^{\beta_1}(z_1,\bar{z}_1) \cdot {g_{\alpha_4}}^{\beta_4}(z_2,\bar{z}_2) \cdot {g^{-1}_{\beta_2}}^{\alpha_2}(z_3,\bar{z}_3) \cdot {g^{-1}_{\beta_3}}^{\alpha_3}(z_4,\bar{z}_4) \rangle
\eea
and their transformation properties under the modular tranformations (after the identification \pref{idef} described in section \ref{blocks}). We will refer to the correlators listed above as the first, second and third correlators. Blocks
and their transformation matrices will be given subscripts to indicate the correlator they belong to.

For the first correlator
$$ \langle  {g_{\alpha_1}}^{\beta_1}(z_1,\bar{z}_1) \cdot {g^{-1}_{\beta_2}}^{\alpha_2}(z_2,\bar{z}_2) \cdot {g^{-1}_{\beta_3}}^{\alpha_3}(z_3,\bar{z}_3) \cdot {g_{\alpha_4}}^{\beta_4}(z_4,\bar{z}_4) \rangle $$
the holomorphic conformal blocks\footnote{The blocks for this correlator have already been discussed in the main text. We rewrite them here with 
the subscript convention discussed above, so as to have a consistent notation for this Appendix.} are
\begin{eqnarray}
\nonumber
  \mathcal{F}^{\mathds{1}}_{(1)}(x) &=& I_1 F^{\mathds{1}}_{(1)1}(x)  + I_2 F^{\mathds{1}}_{(1)2}(x),\\
  \mathcal{F}^{\theta}_{(1)}(x) &=&  I_1 F^{\mathds{\theta}}_{(1)1}(x)  + I_2 F^{\theta}_{(1)2}(x),
\end{eqnarray}
where 
\begin{eqnarray}
\nonumber
 F^{\mathds{1}}_{(1)1}(x) &=& x^{-\frac{4h_g}{3}}(1-x)^{h_{{\theta}}-\frac{4h_g}{3}} F \left(\frac{1}{\tilde{k}},-\frac{1}{\tilde{k}};1-\frac{N}{\tilde{k}};x \right) \ , \\
 \nonumber
 F^{\mathds{1}}_{(1)2}(x)&=&  {1 \over k} x^{1-\frac{4h_g}{3}}(1-x)^{h_{{\theta}}-\frac{4h_g}{3}} F\left(1+\frac{1}{\tilde{k}},1-\frac{1}{\tilde{k}};2-\frac{N}{\tilde{k}};x \right) \ , \\
 \nonumber
 F^{\mathds{\theta}}_{(1)1}(x)&=& x^{h_{{\theta}}-\frac{4h_g}{3}}(1-x)^{h_{{\theta}}-\frac{4h_g}{3}} F\left(\frac{N}{\tilde{k}}-\frac{1}{\tilde{k}},\frac{N}{\tilde{k}}+\frac{1}{\tilde{k}};1+\frac{N}{\tilde{k}};x \right) \ , \\ 
 F^{\mathds{\theta}}_{(1)2} (x) &=& - Nx^{h_{\theta}-\frac{4h_g}{3}}(1-x)^{h_{\theta}-\frac{4h_g}{3}} F \left(\frac{N}{\tilde{k}}-\frac{1}{\tilde{k}},\frac{N}{\tilde{k}}+\frac{1} {\tilde{k}};\frac{N}{\tilde{k}};x \right) \ .
\end{eqnarray}
The holomorphic blocks for the correlator
$$ \langle  {g_{\alpha_1}}^{\beta_1}(z_1,\bar{z}_1) \cdot {g^{-1}_{\beta_2}}^{\alpha_2}(z_2,\bar{z}_2) \cdot {g_{\alpha_4}}^{\beta_4}(z_3,\bar{z}_3) \cdot {g^{-1}_{\beta_3}}^{\alpha_3}(z_4,\bar{z}_4) \rangle
 $$
are
\begin{eqnarray}
\nonumber
  \mathcal{F}^{\mathds{1}}_{(2)}(x) &=& I_1 F^{\mathds{1}}_{(2)1}(x)  + I_2 F^{\mathds{1}}_{(2)2}(x),\\
  \mathcal{F}^{\theta}_{(2)}(x) &=&  I_1 F^{\mathds{\theta}}_{(2)1}(x)  + I_2 F^{\theta}_{(2)2}(x),
\end{eqnarray}
where
\begin{eqnarray}
\nonumber
 F^{\mathds{1}}_{(2)1}(x) &=& x^{-\frac{4h_g}{3}}(1-x)^{h_{\chi}-\frac{4h_g}{3}} F\left(\frac{1}{\tilde{k}},1-\frac{N}{\tilde{k}}+\frac{1}{\tilde{k}};1-\frac{N}{\tilde{k}};x\right) \ , \\
 \nonumber
 F^{\mathds{1}}_{(2)2}(x)&=& -{1 \over k} x^{1-\frac{4h_g}{3}}(1-x)^{h_{\chi}-\frac{4h_g}{3}} F\left(1+\frac{1}{\tilde{k}},1-\frac{N}{\tilde{k}}+\frac{1}{\tilde{k}};2-\frac{N}{\tilde{k}};x\right) \ , \\
 \nonumber
 F^{\mathds{\theta}}_{(2)1}(x) &=& x^{h_{\hat{\theta}}-\frac{4h_g}{3}}(1-x)^{h_{\chi}-\frac{4h_g}{3}} F\left(1+\frac{1}{\tilde{k}},\frac{N}{\tilde{k}}+\frac{1}{\tilde{k}};1+\frac{N}{\tilde{k}};x\right) \ , \\ 
 F^{\mathds{\theta}}_{(2)2} (x) &=& -N x^{h_{\hat{\theta}}-\frac{4h_g}{3}}(1-x)^{h_{\chi}-\frac{4h_g}{3}} F\left(\frac{1}{\tilde{k}},\frac{N}{\tilde{k}}+\frac{1}{\tilde{k}};\frac{N}{\tilde{k}};x\right) \ .
\end{eqnarray}
The holomorphic blocks for the correlator 
$$ \langle  {g_{\alpha_1}}^{\beta_1}(z_1,\bar{z}_1) \cdot {g_{\alpha_4}}^{\beta_4}(z_2,\bar{z}_2) \cdot {g^{-1}_{\beta_2}}^{\alpha_2}(z_3,\bar{z}_3) \cdot {g^{-1}_{\beta_3}}^{\alpha_3}(z_4,\bar{z}_4) \rangle $$
are
\begin{eqnarray}
\nonumber
  \mathcal{F}^{\mathds{\xi}}_{(3)}(x) &=& I_1 F^{\mathds{\xi}}_{(3)1}(x)  + I_2 F^{\mathds{\xi}}_{(3)2}(x),\\
  \mathcal{F}^{\chi}_{(3)}(x) &=&  I_1 F^{\mathds{\chi}}_{(3)1}(x)  + I_2 F^{\chi}_{(3)2}(x),
\end{eqnarray}
where 
\begin{eqnarray}
\nonumber
 F^{\mathds{\xi}}_{(3)1}(x) &=& x^{h_{\xi}-\frac{4h_g}{3}}(1-x)^{h_{\hat{\theta}}-\frac{4h_g}{3}} F\left(1-\frac{1}{\tilde{k}},\frac{N}{\tilde{k}}-\frac{1}{\tilde{k}};1-\frac{2}{\tilde{k}};x\right) \ , \\
 \nonumber
 F^{\mathds{\xi}}_{(3)2}(x)&=& -x^{h_{\xi}-\frac{4h_g}{3}}(1-x)^{h_{\hat{\theta}}-\frac{4h_g}{3}} F\left(-\frac{1}{\tilde{k}},\frac{N}{\tilde{k}}-\frac{1}{\tilde{k}};1-\frac{2}{\tilde{k}};x\right) \ , \\
 \nonumber
 F^{\mathds{\chi}}_{(3)1}(x) &=& x^{h_{\chi}-\frac{4h_g}{3}}(1-x)^{h_{\hat{\theta}}-\frac{4h_g}{3}} F\left(1+\frac{1}{\tilde{k}},\frac{N}{\tilde{k}}+\frac{1}{\tilde{k}};1+\frac{2}{\tilde{k}};x\right) \ , \\ 
 F^{\mathds{\chi}}_{(3)2} (x) &=& x^{h_{\chi}-\frac{4h_g}{3}}(1-x)^{h_{\hat{\theta}}-\frac{4h_g}{3}} F\left(\frac{1}{\tilde{k}},\frac{N}{\tilde{k}}+\frac{1}{\tilde{k}};1+\frac{2}{\tilde{k}};x\right) \ .
\end{eqnarray}
With the  choices for tensor  indices as in \pref{idef},  we will denote the   holomorphic blocks of the three correlators 
by  $\mathcal{H}^{i}_{(q)}(x)$ with $q=1,2,3$  i.e.
\begin{eqnarray}
\nonumber
  \mathcal{H}^{\mathds{1}}_{(1)}(x) &=& I F^{\mathds{1}}_{(1)1}(x)  + I F^{\mathds{1}}_{(1)2}(x),\\
  \nonumber
  \mathcal{H}^{\theta}_{(1)}(x) &=&  I F^{\mathds{\theta}}_{(1)1}(x)  + I F^{\theta}_{(1)2}(x),\\
  \nonumber
  \mathcal{H}^{\mathds{1}}_{(2)}(x) &=& I F^{\mathds{1}}_{(2)1}(x)  + I F^{\mathds{1}}_{(2)2}(x),\\
  \nonumber
  \mathcal{H}^{\theta}_{(2)}(x) &=&  I F^{\mathds{\theta}}_{(2)1}(x)  + I F^{\theta}_{(2)2}(x),\\
  \nonumber
  \mathcal{H}^{\mathds{\xi}}_{(3)}(x) &=& I F^{\mathds{\xi}}_{(3)1}(x)  + I F^{\mathds{\xi}}_{(3)2}(x),\\
\label{six}
  \mathcal{H}^{\chi}_{(3)}(x) &=&  I F^{\mathds{\chi}}_{(3)1}(x)  + I F^{\chi}_{(3)2}(x).
\end{eqnarray}
We note that with $I_1 = I_2$ the three correlators are equal to those in \pref{modvec}.  

The actions of $T$ and $S$ on these can be computed using the following identities of hypergeometric functions~\cite{Weisstein2}.
\begin{equation}
  \begin{aligned}
    F(a,b;c;z)=&(1-z)^{c-a-b}F(c-a,c-b;c;z),\\
    F(a,b;c;\frac{z}{z-1})=&(1-z)^{a}F(a,c-b;c;z)=(1-z)^{b}F(c-a,b;c;z),\\
    F(a,b;c;1-z)=&\frac{\Gamma(c)\Gamma(c-a-b)}{\Gamma(c-a)\Gamma(c-b)}F(a,b;a+b-c+1;z)\\
    &+\frac{\Gamma(c)\Gamma(a+b-c)}{\Gamma(a)\Gamma(b)}z^{c-a-b}F(c-a,c-b;c-a-b+1;z).
  \end{aligned}
\end{equation}
   \begin{eqnarray}
\label{saction}
\nonumber
 F(a,b;c;1-z)=
    \frac{\Gamma(c)\Gamma(a+b-c)}{\Gamma(a)\Gamma(b)}z^{c-a-b} (1 - z)^{1-c}F(1-b,1-a; 1+ c-a-b, z) \\
    +\frac{\Gamma(c)\Gamma(c-a-b)}{\Gamma(c-a)\Gamma(c-b)} (1-z)^{1-c}F(1+ b-c,1+a -c; 1+a+b-c;z)
 \end{eqnarray}
\noindent \underline{Action of $T$}: The action of $T$ on the blocks $\mathcal{H}^{i}_{(1)}(x)$ are given by
\begin{equation}
 \mathcal{H}^{i}_{(1)} \left(T.x \right) = \mathcal{H}^{j}_{(2)} \left(x \right) M_{(1)ji}(T),
\end{equation}
where
\begin{equation}
 M_{(1)}(T)=(-1)^{-2(N^2-1)/3N\tilde{k}}
 \begin{pmatrix}
    1 & 0\\
    0 & (-1)^{N/\tilde{k}}
 \end{pmatrix}.
\end{equation}
The action of $T$ on the blocks $\mathcal{H}^{i}_{(2)}(x)$ are given by
\begin{equation}
 \mathcal{H}^{i}_{(2)} \left(T.x \right) = \mathcal{H}^{j}_{(1)} \left(x \right) M_{(2)ji}(T),
\end{equation}
where
\begin{equation}
 M_{(2)}(T)=(-1)^{-2(N^2-1)/3N\tilde{k}}
 \begin{pmatrix}
    1 & 0\\
    0 & (-1)^{N/\tilde{k}}
 \end{pmatrix}.
\end{equation}
The action of $T$ on the blocks $\mathcal{H}^{i}_{(3)}(x)$ are given by
\begin{equation}
 \mathcal{H}^{i}_{(3)} \left(T.x \right) = \mathcal{H}^{j}_{(3)} \left(x \right) M_{(3)ji}(T),
\end{equation}
where
\begin{equation}
 M_{(3)}(T)=-(-1)^{(N^2-3N-4)/3N\tilde{k}}
 \begin{pmatrix}
    1 & 0\\
    0 & -(-1)^{2/\tilde{k}}
 \end{pmatrix}.
\end{equation}
\noindent \underline{Action of $S$}: The action of $S$ on the blocks $\mathcal{H}^{i}_{(1)}(x)$ are given by
\begin{equation}
 \mathcal{H}^{i}_{(1)} \left(S.x \right) = \mathcal{H}^{j}_{(1)} \left(x \right) M_{(1)ji}(S),
\end{equation}
where
\begin{equation}
 M_{(1)}(S)=
  \begin{pmatrix}
    -\frac{\tilde{k}\Gamma \left({N} {/}{\tilde{k}} \right)\Gamma \left({k}{/}{\tilde{k}}\right)}{\Gamma \left({1}{/}{\tilde{k}}\right)\Gamma \left(-{1}{/}{\tilde{k}}\right)} & -\frac{N\Gamma^2 \left({N}{/}{\tilde{k}}\right)}{\Gamma \left({N}{/}{\tilde{k}}-{1}{/}{\tilde{k}}\right)\Gamma \left({N}{/}{\tilde{k}}+{1}{/}{\tilde{k}}\right)}\\
    -\frac{\Gamma^2 \left({k}{/}{\tilde{k}}\right)}{N\Gamma \left(k/ \tilde{k}  - 1/ \tilde{k} \right)\Gamma \left(k/ \tilde{k}  + 1/ \tilde{k} \right)} & \frac{\tilde{k}\Gamma \left({N}{/}{\tilde{k}}\right)\Gamma \left({k}{/}{\tilde{k}}\right)}{\Gamma \left({1}{/}{\tilde{k}}\right)\Gamma \left(-{1}{/}{\tilde{k}}\right)}
  \end{pmatrix}.
\end{equation}
The action of $S$ on the blocks $\mathcal{H}^{i}_{(2)}(x)$ are given by
\begin{equation}
 \mathcal{H}^{i}_{(2)} \left(S.x \right) = \mathcal{H}^{j}_{(3)} \left(x \right) M_{(2)ji}(S),
\end{equation}
where
\begin{equation}
 M_{(2)}(S)=
  \begin{pmatrix}
    \frac{\Gamma\left(k/\tilde{k}\right)\Gamma\left(2/\tilde{k}\right)}{\Gamma\left(1/\tilde{k}\right)\Gamma\left(k/\tilde{k}+1/\tilde{k}\right)} & \frac{N\Gamma\left(N/\tilde{k}\right)\Gamma\left(2/\tilde{k}\right)}{\Gamma\left(1/\tilde{k}\right)\Gamma\left(N/\tilde{k}+1/\tilde{k}\right)}\\
    \frac{\Gamma\left(k/\tilde{k}\right)\Gamma\left(-2/\tilde{k}\right)}{\Gamma\left(k/\tilde{k}-1/\tilde{k}\right)\Gamma\left(-1/\tilde{k}\right)} & -\frac{N\Gamma\left(N/\tilde{k}\right)\Gamma\left(-2/\tilde{k}\right)}{\Gamma\left(N/\tilde{k}-1/\tilde{k}\right)\Gamma\left(-1/\tilde{k}\right)}
  \end{pmatrix}.
\end{equation}
The action of $S$ on the blocks $\mathcal{H}^{i}_{(3)}(x)$ are given by
\begin{equation}
 \mathcal{H}^{i}_{(3)} \left(S.x \right) = \mathcal{H}^{j}_{(2)} \left(x \right) M_{(3)ji}(S),
\end{equation}
where
\begin{equation}
 M_{(3)}(S)=
  \begin{pmatrix}
    \frac{2\Gamma\left(-2/\tilde{k}\right)\Gamma\left(N/\tilde{k}\right)}{ \Gamma\left(-1/\tilde{k}\right) \Gamma\left(N/\tilde{k}-1/\tilde{k}\right)} & \frac{2\Gamma\left(2/\tilde{k}\right)\Gamma\left(N/\tilde{k}\right)}{ \Gamma\left(1/\tilde{k}\right) \Gamma\left(N/\tilde{k}+1/\tilde{k}\right)}\\
    \frac{\Gamma\left(1-2/\tilde{k}\right)\Gamma\left(-N/\tilde{k}\right)}{ \Gamma\left(-1/\tilde{k}\right) \Gamma\left(k/\tilde{k}-1/\tilde{k}\right)} & \frac{\Gamma\left(1+2/\tilde{k}\right)\Gamma\left(-N/\tilde{k}\right)}{ \Gamma\left(1/\tilde{k}\right) \Gamma\left(k/\tilde{k}+1/\tilde{k}\right)}
  \end{pmatrix}.
\end{equation}

\section{Generators of the orbit for $N=k$ theories}
\label{neqk}

In this section, we show that   for general values of $N(=k)$ the orbit of $C_{\rm seed}$ is  as given in \pref{orbneqk}. We will do this by showing that the orbit can in effect
be generated by considering the action of matrices of the form
\begin{equation}
\label{genorbneqk}
\begin{pmatrix} 
\sin \alpha & -k\cos\alpha \\ -\frac{1}{k}\cos\alpha & -\sin\alpha 
\end{pmatrix},
\end{equation}
on $C_{\rm seed}$, where $ \alpha = {\pi (2s +1)\over 2 k} $ with $s = 0 \cdots (k-1)$ for $k$ odd, and  $ \alpha = {\pi s\over 2k}$ with $s = 0 \cdots (2k-1)$ for $k$ even.
It is easy to check that the actions of these matrices on $C_{\rm seed}$ indeed generates the orbits described in  \pref{orbneqk}.
  We begin by noting that for $M(\gamma)$ of the form
$$M(\gamma)\equiv\begin{pmatrix} a_{\gamma} & b_{\gamma}\\c_{\gamma} & d_{\gamma} \end{pmatrix},$$
its action on $C_{\rm seed}$ yields
\begin{equation}
\label{orbbb}
  \begin{pmatrix} |a_{\gamma}|^2 & a_{\gamma}c_{\gamma}^{*} \\ a_{\gamma}^{*}c_{\gamma} & |c_{\gamma}|^2 \end{pmatrix}.
\end{equation}
Thus, the result of the action   only  depends on $a_{\gamma}$ and $c_{\gamma}$ (and is independent of $b_{\gamma}$ and $d_{\gamma}$). Furthermore, since \pref{orbbb}
is quadratic in $a_{\gamma}$ and $c_{\gamma}$, elements of the orbit are only sensitive to their relative sign. Thus   deformations of
$M(\gamma)s$ which modify $b_{\gamma}$, $d_{\gamma}$ and the relative sign between $a_{\gamma}$, $c_{\gamma}$ keep their actions on $C_{\rm seed}$  unchanged. We will use
such deformations to  show that the orbit is in effect generated by the matrices given in \pref{genorbneqk}.  Let us start by considering the first few matrices in the list \pref{mlist} of $M(\gamma)$ (for theories with $N=k$). In what follows, we will use the symbol `$\sim$' to denote a deformation of a matrix $M(\gamma)$ which keeps its action on $C_{\rm seed}$ unchanged.
\begin{flalign}
\nonumber
&\phantom{abcde} M(\mathds{1})=\begin{pmatrix} 1 & 0 \\ 0 & 1 \end{pmatrix} \sim \begin{pmatrix} 1 & 0 \\ 0 & -1 \end{pmatrix}=\begin{pmatrix} \sin\frac{\pi k}{2k} & -k\cos\frac{\pi k}{2k} \\ -\frac{1}{k}\cos\frac{\pi k}{2k} & -\sin\frac{\pi k}{2k} \end{pmatrix};&
\end{flalign}
\begin{flalign}
\nonumber
&\phantom{abcde} M(S)=\begin{pmatrix} \sin\frac{\pi}{2k} & -k\cos\frac{\pi}{2k}\\ -\frac{1}{k}\cos\frac{\pi}{2k} & -\sin\frac{\pi}{2k} \end{pmatrix};&
\end{flalign}
\begin{flalign}
\nonumber
&\phantom{abcde} M(ST^2)=\begin{pmatrix} \sin\frac{\pi}{2k} & -k\cos\frac{\pi}{2k}\\ \frac{1}{k}\cos\frac{\pi}{2k} & \sin\frac{\pi}{2k} \end{pmatrix} \sim \begin{pmatrix} \sin\frac{\pi(2k-1)}{2k} & -k\cos\frac{\pi(2k-1)}{2k}\\ -\frac{1}{k}\cos\frac{\pi(2k-1)}{2k} & -\sin\frac{\pi(2k-1)}{2k} \end{pmatrix};&
\end{flalign}
\begin{flalign}
\nonumber
&\phantom{abcde} M(ST^2S)= \begin{pmatrix} \sin\frac{\pi(2-k)}{2k} & -k\cos\frac{\pi(2-k)}{2k}\\ -\frac{1}{k}\cos\frac{\pi(2-k)}{2k} & -\sin\frac{\pi(2-k)}{2k} \end{pmatrix} \sim \begin{pmatrix} \sin\frac{\pi(2+k)}{2k} & -k\cos\frac{\pi(2+k)}{2k}\\ -\frac{1}{k}\cos\frac{\pi(2+k)}{2k} & -\sin\frac{\pi(2+k)}{2k} \end{pmatrix};&
\end{flalign}
\begin{flalign}
\nonumber
&\phantom{abcde} \begin{matrix}
M(ST^2ST^2) = \begin{pmatrix} -\cos\frac{2\pi}{2k} & -k\sin\frac{2\pi}{2k}\\ \frac{1}{k}\sin\frac{2\pi}{2k} & -\cos\frac{2\pi}{2k} \end{pmatrix} \sim \begin{pmatrix} \sin\frac{\pi(3k-2)}{2k} & -k\cos\frac{\pi(3k-2)}{2k}\\ -\frac{1}{k}\cos\frac{\pi(3k-2)}{2k} & -\sin\frac{\pi(3k-2)}{2k} \end{pmatrix}\\
\phantom{abcdefg}\sim \begin{pmatrix} \sin\frac{\pi(k-2)}{2k} & -k\cos\frac{\pi(k-2)}{2k}\\ -\frac{1}{k}\cos\frac{\pi(k-2)}{2k} & -\sin\frac{\pi(k-2)}{2k} \end{pmatrix};
\end{matrix}&
\end{flalign}
\begin{flalign}
\nonumber
&\phantom{abcde} M(ST^2ST^2S)=\begin{pmatrix} -\sin\frac{3\pi}{2k} & k\cos\frac{3\pi}{2k}\\ \frac{1}{k}\cos\frac{3\pi}{2k} & \sin\frac{3\pi}{2k} \end{pmatrix} \sim \begin{pmatrix} \sin\frac{3\pi}{2k} & -k\cos\frac{3\pi}{2k}\\ -\frac{1}{k}\cos\frac{3\pi}{2k} & -\sin\frac{3\pi}{2k} \end{pmatrix}.&
\end{flalign}
Proceeding as above, all the $M(\gamma)$ can be brought to the form in \pref{genorbneqk} by making use of the identities
$$
\begin{pmatrix} \sin\beta & -k\cos\beta\\ -\frac{1}{k}\cos\beta & -\sin\beta \end{pmatrix} . \begin{pmatrix} 1 & 0\\ 0 & -1 \end{pmatrix} . \begin{pmatrix} \sin\alpha & -k\cos\alpha\\ -\frac{1}{k}\cos\alpha & -\sin\alpha \end{pmatrix}= \begin{pmatrix} \sin(\alpha+\beta-\frac{\pi}{2}) & -k\cos(\alpha+\beta-\frac{\pi}{2}) \\ -\frac{1}{k}\cos(\alpha+\beta-\frac{\pi}{2}) & -\sin(\alpha+\beta-\frac{\pi}{2}) \end{pmatrix}
$$
and 
$$
\begin{pmatrix} 
\sin \alpha & -k\cos\alpha \\ -\frac{1}{k}\cos\alpha & -\sin\alpha 
\end{pmatrix}
\sim
\begin{pmatrix} 
\sin \left(\alpha + \pi \right) & -k\cos \left(\alpha + \pi \right) \\ -\frac{1}{k}\cos  \left(\alpha + \pi \right)  & -\sin  \left(\alpha + \pi \right)
\end{pmatrix}
$$ 
for any angle $\alpha$ and $\beta$.

For completeness, we provide the orbit the $N(=k)=2$ theory. It can easily be checked that this is same as that given by  the matrices in \pref{orbneqk}.
 For $N=2, k=2$ the matrices $M(S)$ and $M(T^2)$ are
\begin{equation}
M(S)=\begin{pmatrix} \frac{1}{\sqrt{2}} & -\sqrt{2} \\ -\frac{1}{2\sqrt{2}} & -\frac{1}{\sqrt{2}} \end{pmatrix},  \phantom{abcd}  M(T^2)= e^{- \frac{i\pi }{2}} \begin{pmatrix} 1 & 0\\ 0 & -1 \end{pmatrix}.
\end{equation}
The orbit of $C_{\rm seed}$ consists of four matrices. It is generated by the action of $\mathds{1}, S, ST^2$ and $ST^2S$. We tabulate the results of these actions
in  Table~\ref{tab:2}. The normalised sum over the orbit \pref{ourav} reproduces the KZ result.

\begin{table}[H]
\begin{center}
  \begin{tabular}{|c|c|}
    \hline
    $\gamma$ & $\text{ } M(\gamma) \cdot C_{\rm seed} \cdot M(\gamma)^{\dagger} \text{ }$\\
    \hline
    $\mathds{1}$ & $\begin{pmatrix} 1 & 0 \\ 0 & 0 \end{pmatrix}$\\[0.4cm]
    $S$ & $\begin{pmatrix} \frac{1}{2} & -\frac{1}{4} \\ -\frac{1}{4} & \frac{1}{8} \end{pmatrix}$\\[0.4cm]
    $ST^2$ & $\begin{pmatrix} \frac{1}{2} & \frac{1}{4} \\ \frac{1}{4} & \frac{1}{8} \end{pmatrix}$\\[0.4cm]
    $ST^2S$ & $\begin{pmatrix} 0 & 0 \\ 0 & \frac{1}{4} \end{pmatrix}$\\[0.4cm]
    \hline
    $X^{\rm av}$  & $\begin{pmatrix} 1 & 0 \\ 0 & \frac{1}{4} \end{pmatrix}$\\[0.4cm]
    \hline
  \end{tabular}
\end{center}
\caption{ Orbit of the vacuum block  for  $N=2, k=2$}
\label{tab:2}
\end{table}

\section{Further numerical examples}
\label{exnneqk}

Here we provide a couple of examples where the numerics are quite involved as discussed at the end of section \ref{Saverage}.

\begin{figure}[H]
  \centering
    \includegraphics[width=8.0cm]{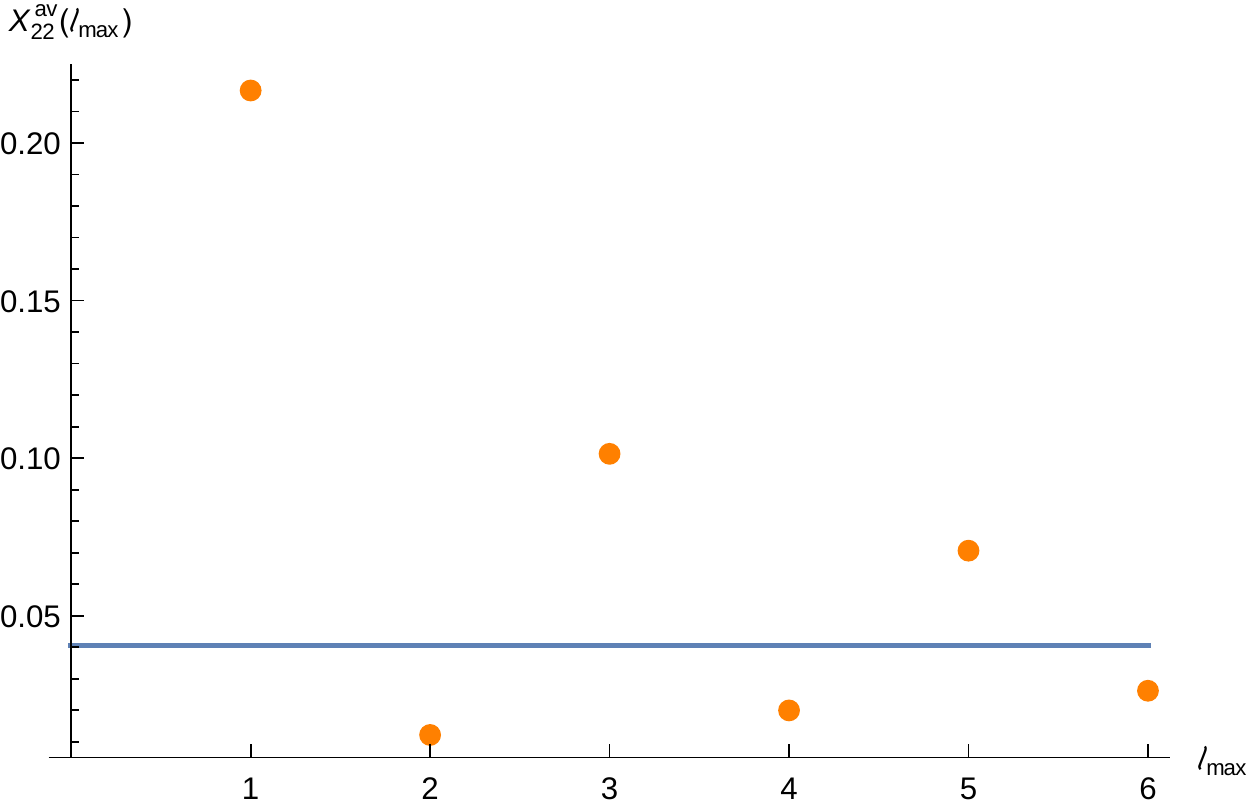}
  \caption{Orange dots show $X^{\rm av}_{22}(\ell_{\rm max})$ in the range $[0.005, 0.225]$ plotted against $\ell_{\rm max}$. Blue horizontal line at $0.0405346$ represents $X^{\rm KZ}_{22}$.}
\label{fig:5}
\end{figure}

\noindent \underline{ $N=5$, $k=6$:} For $N=5, k =6$, the value of $m(5,6)$ as defined in \pref{mdef} is 11.  Thus with each increment in $\ell_{\rm max}$ by $1$, there is approximately a tenfold increase in the number of new terms added to the sum \pref{numereq}. With the available computing resources we have performed the sum upto $\ell_{\rm max}=6$. This involves $1193006$ distinct contributions to the sum. We find $X^{\rm av}_{22}(6) = 0.026177$, alongside we note the exact result  \pref{kzr}, $X^{\rm KZ}_{22} \approx 0.0405346$. The off diagonal entries of $X^{\rm av}(6)$ are of the order of $10^{-14}$. Figure ~\ref{fig:5} shows our results for $X^{\rm av}_{22}(\ell_{\rm max})$ as a function of $\ell_{\rm max}$,  all qualitative features of the numerics are same as those in the  examples discussed in section \ref{Saverage}.

\begin{figure}[H]
  \centering
    \includegraphics[width=8.0cm]{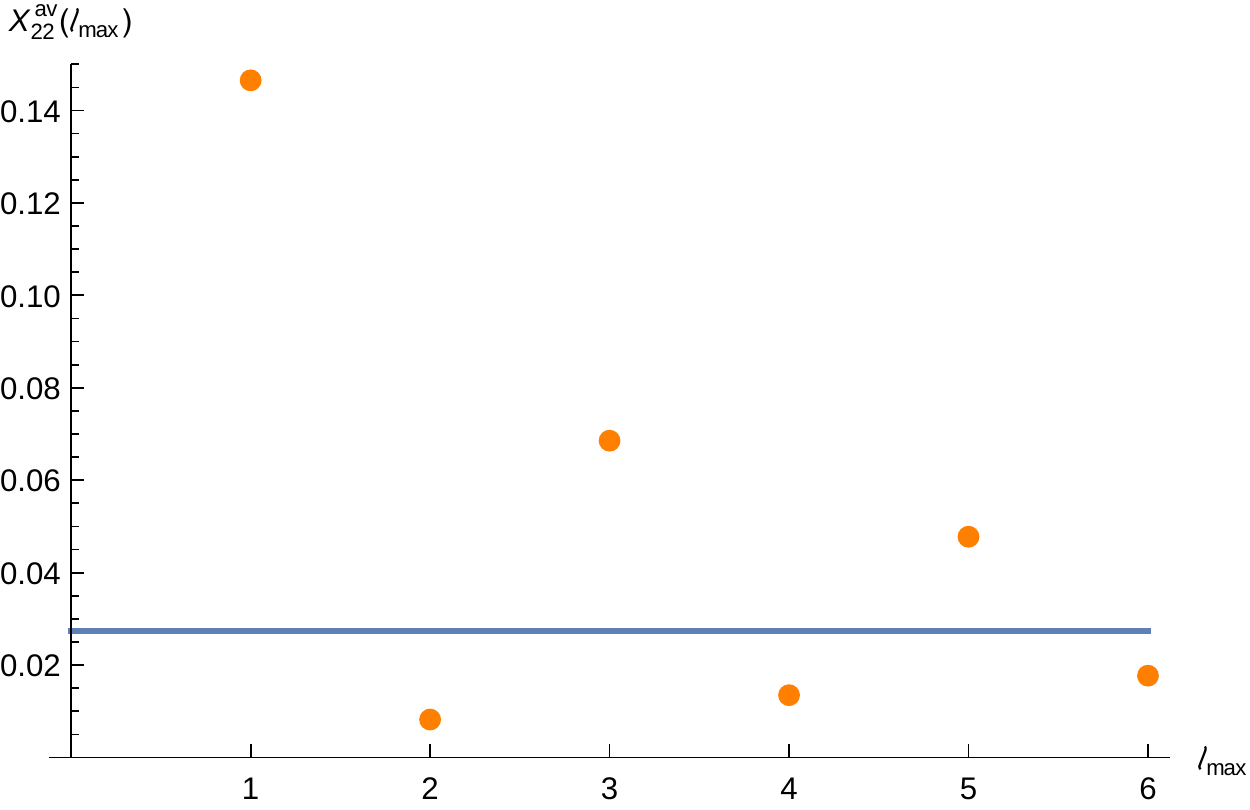}
  \caption{Orange dots show $X^{\rm av}_{22}(\ell_{\rm max})$ in the range $[0.000, 0.150]$ plotted against $\ell_{\rm max}$. Blue horizontal line at $0.0274114$ represents $X^{\rm KZ}_{22}$.}
\label{fig:6}
\end{figure}

\noindent \underline{ $N=6$, $k=5$:} For $N=6, k =5$, the value of $m(6,5)$ as defined in \pref{mdef} is 11.  Thus similarly, with each increment in $\ell_{\rm max}$ by $1$, there is approximately a tenfold increase in the number of new terms added to the sum \pref{numereq}. With the available computing resources we have performed the sum upto $\ell_{\rm max}=6$. This involves $1193006$ distinct contributions to the sum. We find  $X^{\rm av}_{22}(6) = 0.0177022$, alongside we note the exact result  \pref{kzr}, $X^{\rm KZ}_{22} \approx 0.0274114$. The off diagonal entries of $X^{\rm av}(6)$ are of the order of $10^{-14}$. Figure ~\ref{fig:6} shows our results for $X^{\rm av}_{22}(\ell_{\rm max})$ as a function of $\ell_{\rm max}$. All the features of the numerics are similar to the previous example.

\section{Averaging over all of $PSL(2, \mathbb{Z})$ }
\label{apmodav}

In this Appendix, we briefly discuss the construction of correlator from averaging over the full modular group. To implement the prescription \pref{aver},
the six holomorphic blocks  in \pref{six} of the three correlators in \pref{modvec} can be put in a six dimensional row:
\begin{equation}
\vec{\mathcal{H}}(\tau)=\bigg(\mathcal{H}^{\mathds{1}}_{(1)}(\tau), \mathcal{H}^{\theta}_{(1)}(\tau), \mathcal{H}^{\mathds{1}}_{(2)}(\tau), \mathcal{H}^{\theta}_{(2)}(\tau), \mathcal{H}^{\mathds{\xi}}_{(3)}(\tau), \mathcal{H}^{\chi}_{(3)}(\tau)\bigg)\ .
\end{equation}
On this, $T$ and $S$ act as
\begin{equation}
\mathcal{H}^{i}(T.\tau) = \mathcal{H}^j(\tau)\mathcal{M}_{ji}(T) \phantom{a} {\rm and}\phantom{a} \mathcal{H}^{i}(S.\tau) = \mathcal{H}^j(\tau)\mathcal{M}_{ji}(S)\ 
\end{equation}
with
\begin{equation}
\mathcal{M}(T)=
\begin{pmatrix}
  0 & M_{(1)}(T) & 0\\
  M_{(2)}(T) & 0 & 0\\
  0 & 0 & M_{(3)}(T)
\end{pmatrix}
\phantom{a} {\rm and} \phantom{b}
\mathcal{M}(S)=
\begin{pmatrix}
  M_{(1)}(S) & 0 & 0\\
  0 & 0 & M_{(2)}(S)\\
  0 & M_{(3)}(S) & 0
\end{pmatrix},
\end{equation}
where the two dimensional matrices ($M_{(i)}(T)$ and $M_{(i)}(S)$) are as defined in Appendix \ref{apblocks}. The light contribution as defined in \pref{gencor2} 
can be taken as
\begin{equation}
G^{\rm light}_B(\tau,\bar{\tau})=C^B_{i(B)j(B)} \mathcal{H}^{i(B)}(\tau) \bar{\mathcal{H}}^{j(B)}(\bar{\tau}), \phantom{a} B=1,2,3\ ,
\end{equation}
where repeated indices are summed over with $i(1),j(1)\in\{1,2\}$, $i(2),j(2)\in\{3,4\}$ and $i(3),j(3)\in\{5,6\}$,
\begin{equation}
C^B=\begin{pmatrix} 1 & 0\\0 & 0 \end{pmatrix}, \phantom{a} B=1,2,3\ .
\end{equation}
Under the action  $\gamma\in PSL(2,\mathbb{Z})$,
\begin{equation}
C^B_{i(B)j(B)} \mathcal{H}^{i(B)}(\tau) \bar{\mathcal{H}}^{j(B)}(\bar{\tau}) \to \mathcal{M}(\gamma)_{k i(B)} C^B_{i(B)j(B)} \mathcal{M}(\gamma)^{\dagger}_{j(B)l} \mathcal{H}^{k}(\tau) \bar{\mathcal{H}}^{l}(\bar{\tau})\ .
\end{equation}
For each $\gamma$ we arrange the three $6\times6$ matrices
\begin{equation}
\sigma^{-1}(\gamma)_{AB}\mathcal{M}(\gamma)_{k i(B)} C^B_{i(B)j(B)} \mathcal{M}(\gamma)^{\dagger}_{j(B)l} \phantom{a}, \phantom{a}A=1,2,3\ ,
\end{equation}
in a three dimensional column $\vec{X}(\gamma)$.  The sum \pref{aver} then  reads 
\begin{equation}
\label{fullmodav}
\vec{X}^{\rm av} = \mathcal{N}^{-1} \cdot \sum_{\gamma \in PSL(2,\mathbb{Z})}\vec{X}(\gamma)\ ,
\end{equation}
where the normalisation $\mathcal{N}$ is the $(1,1)$ element of $\big[\sum_{\gamma}\vec{X}(\gamma)\big]^{1}$. Hence the candidate for the vector-valued modular function \pref{modvec} is given by
\begin{equation}
\big[\vec{X}^{\rm av}\big]^{A}_{kl}\mathcal{H}^{k}(\tau) \bar{\mathcal{H}}^{l}(\bar{\tau}), \phantom{a} A=1,2,3\ .
\end{equation}
To incorporate the distinct contributions $\vec{X}(\gamma)$ to the sum \pref{fullmodav}, elements $\gamma$ are arranged in a list similar to \pref{list} where we replace all $T^{2r_i}$ by $T^{r_i}$, and $m$ denotes the smallest positive integer such that
$$\mathcal{M}(T^m) \propto \mathds{1}\ .$$
We perform the sum \pref{fullmodav} taking distinct contributions of elements $\gamma$ of all lengths upto a maximum value $\ell_{\rm max}$:
\begin{equation}
\label{numfulmodav}
\vec{X}^{\rm av}(\ell_{\rm max})=\mathcal{N}(\ell_{\rm max})^{-1} \cdot \sum'_{\ell(\gamma) \leq \ell_{\rm max}} \vec{X}(\gamma)\ ,
\end{equation}
where the primed sum indicates that distinct elements are added. Our results are as follows

\noindent \underline{$N=2$, $k=2$:} For $N=2, k=2$, the sum \pref{numfulmodav} is finite and consists of six distinct contributions, reproducing the KZ result, $\big[\vec{X}^{\rm av}\big]^{1}_{22} = {1 \over 4}$.

\noindent \underline{$N=2$, $k=4$:} For $N=2, k=4$, the sum \pref{numfulmodav} is finite and consists of four distinct contributions, reproducing the KZ result, $\big[\vec{X}^{\rm av}\big]^{1}_{22} = \frac{1}{2 \sqrt[3]{4}}$.

\begin{figure}[H]
  \centering
    \includegraphics[width=8.0cm]{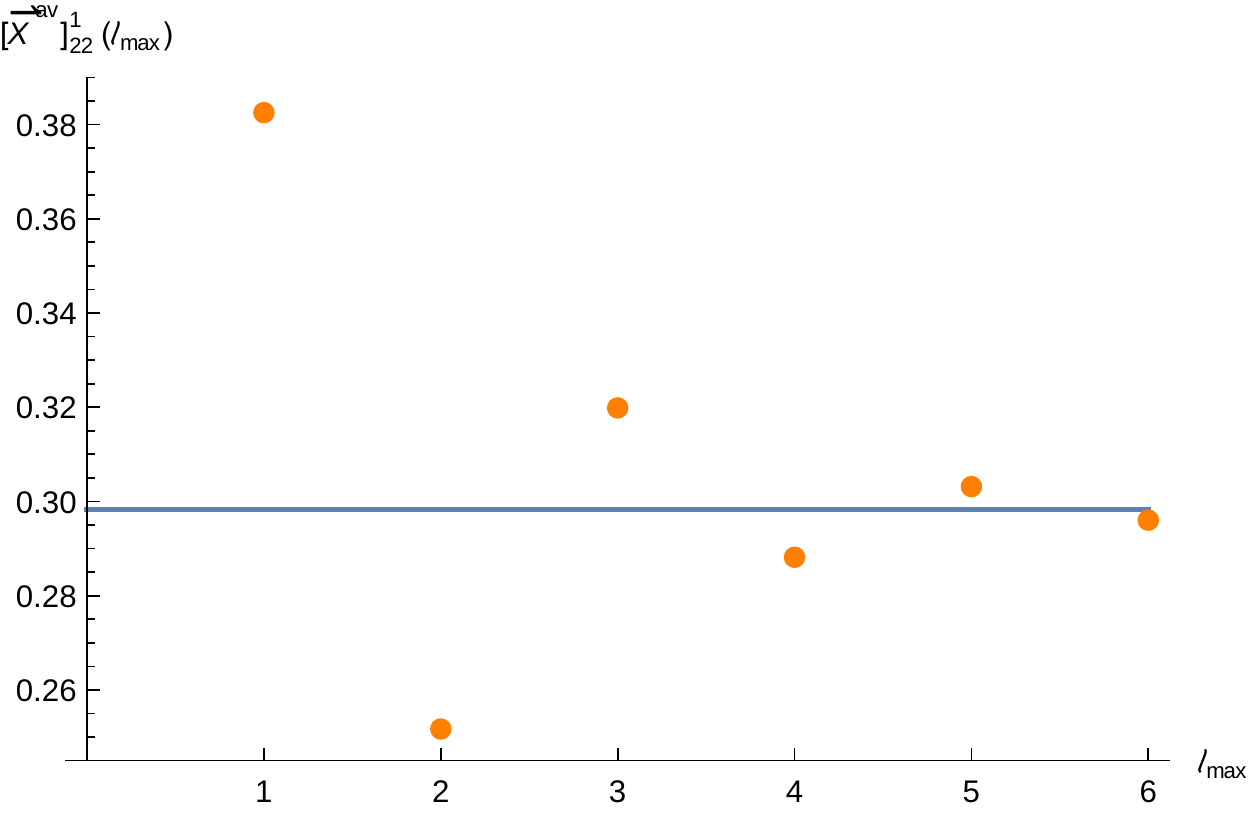}
  \caption{Orange dots show $\big[\vec{X}^{\rm av}\big]^{1}_{22}(\ell_{\rm max})$ in the range $[0.245, 0.390]$ plotted against $\ell_{\rm max}$. Blue horizontal line at $0.29831$ represents the KZ result.}
\label{fig:7}
\end{figure}

\noindent \underline{$N=2$, $k=3$:} For $N=2, k=3$, the sum \pref{numfulmodav} seems to be infinite. We have performed the sum upto $\ell_{\rm max}=6$. This invloves $83651$ distinct contributions to the sum. We find $\big[\vec{X}^{\rm av}\big]^{1}_{22}(6)=0.296026$, which is in good agreement with the KZ result. Figure ~\ref{fig:7} shows our results for $\big[\vec{X}^{\rm av}\big]^{1}_{22}(\ell_{\rm max})$ as a function of $\ell_{\rm max}$.

Increasing $N$ and $k$ makes the numerics quite involved, we leave this for future work.

 \section{The matrices $M^{\ell}_{N,k}$ and $\tilde{M}^{\ell}_{N,k}$}
 \label{matap}
 
    In this Appendix, we obtain the general form of the matrices $M^{\ell}_{N,k}$ and $\tilde{M}^{\ell}_{N,k}$. We then use these to derive the
relations given in \pref{exrelm}. The elements of matrices $M^{\ell}_{N,k}$ can be computed recursively in $\ell$ using their defining equation in \pref{mmdef}
\begin{equation}
  M^{\ell +1}_{N,k}(r_1, \cdots r_{\ell +1}) = M(T^{2r_{\ell+1}})M(S)M^{\ell}_{N,k} (r_1, \cdots r_{\ell}).
\end{equation}
This gives the following relations for the functions that appear in \pref{mpar}
\begin{eqnarray}
\nonumber
a^{\ell +1}_{N,k} (r_1, \cdots r_\ell+1) &=& a_s (N,k)a^{\ell}_{N,k} (r_1 \cdots r_\ell) + b_s(N,k) c_s (N,k)c^{\ell}_{N,k} (r_1 \cdots r_\ell)\\
\nonumber
b^{\ell +1}_{N,k} (r_1, \cdots r_\ell+1)&=& a_s(N,k) b_{N,k}^{\ell}(r_1 \cdots r_\ell) + d_{N,k}^\ell (r_1 \cdots r_\ell)\\
\nonumber
c^{\ell +1}_{N,k} (r_1, \cdots r_\ell+1)&=& e^{i r_{\ell +1}\phi(N,k)} \left( d_s (N,k) c_{N,k}^{\ell} (r_1 \cdots r_\ell)+ a_{N,k}^{\ell} (r_1 \cdots r_\ell)  \right) \\
\nonumber
d^{\ell + 1}_{N,k} (r_1, \cdots r_\ell+1) &=& e^{i r_{\ell +1} \phi(N,k)} \left(  d_s (N,k) d_{N,k}^{\ell} (r_1 \cdots r_\ell) + b_s(N,k) c_s (N,k)b_{N,k}^{\ell} (r_1 \cdots r_\ell) \right)
\end{eqnarray}

Similarly, the matrices  $\tilde{M}^{\ell}_{N,k}$ can be computed recursively in $\ell$ using their defining equation n \pref{mmtdef}
\begin{equation}
   \tilde{M}^{\ell +1}_{N,k}(r_1, \cdots r_{\ell +1}) = M(T^{-2r_{\ell+1}})M(S) \tilde{M}^{\ell}_{N,k} (r_1, \cdots r_{\ell}).
\end{equation}
This gives following relations for the functions that appear in \pref{tmpar}
\begin{eqnarray}
\nonumber
\tilde{a}^{\ell +1}_{N,k} (r_1, \cdots r_\ell+1) &=& a_s (N,k) \tilde{a}^{\ell}_{N,k} (r_1 \cdots r_\ell) + b_s(N,k) c_s (N,k)\tilde{c}^{\ell}_{N,k} (r_1 \cdots r_\ell)\\
\nonumber
\tilde{b}^{\ell +1}_{N,k} (r_1, \cdots r_\ell+1)&=& a_s(N,k) \tilde{b}_{N,k}^{\ell}(r_1 \cdots r_\ell) + \tilde{d}_{N,k}^\ell (r_1 \cdots r_\ell)\\
\nonumber
\tilde{c}^{\ell +1}_{N,k} (r_1, \cdots r_\ell+1)&=& e^{-i r_{\ell +1}\phi(N,k)} \left( d_s (N,k) \tilde{c}_{N,k}^{\ell} (r_1 \cdots r_\ell)+ \tilde{a}_{N,k}^{\ell} (r_1 \cdots r_\ell)  \right) \\
\nonumber
\tilde{d}^{\ell + 1}_{N,k} (r_1, \cdots r_\ell+1) &=& e^{-i r_{\ell +1} \phi(N,k)} \left(  d_s (N,k) \tilde{d}_{N,k}^{\ell} (r_1 \cdots r_\ell) + b_s(N,k) c_s (N,k)\tilde{b}_{N,k}^{\ell} (r_1 \cdots r_\ell) \right).
\end{eqnarray}
Now, making use of relations in \pref{srel} and the fact that\footnote{Recall that $\phi(N,k) = { 2 \pi N \over {k +N}}$.}
\begin{equation}
 e^{i r \phi(N,k)} =  e^{- i r \phi(k,N)} \ \ \text{for any integer r},
\end{equation}
it is easy to see that $\tilde{a}^{\ell}_{k,N}(r_i), \tilde{b}^{\ell}_{k,N}(r_i), \tilde{c}^{\ell}_{k,N}(r_i), \tilde{d}^{\ell}_{k,N}(r_i)$ have exactly the same recurrence relations as $a^{\ell}_{N,k}(r_i),  b^{\ell}_{N,k}(r_i), c^{\ell}_{N,k}(r_i), d^{\ell}_{N,k}(r_i)$. Given that they have same initial values, hence the equalities in \pref{exrelm}.

\providecommand{\href}[2]{#2}\begingroup\raggedright
\endgroup
\end{document}